\documentclass[pdflatex,sn-mathphys-num]{sn-jnl}

\usepackage{geometry}
\usepackage{graphicx}
\usepackage{dcolumn}
\usepackage{bm}
\usepackage{amsfonts}
\usepackage{amsmath}
\usepackage{appendix}
\usepackage{orcidlink}
\usepackage{tikz}
\usetikzlibrary{tikzmark}

\usepackage{multirow}%
\usepackage{amsmath,amssymb,amsfonts}%
\usepackage{amsthm}%
\usepackage{mathrsfs}%
\usepackage[title]{appendix}%
\usepackage{xcolor}%
\usepackage{textcomp}%
\usepackage{manyfoot}%
\usepackage{booktabs}%
\usepackage{algorithm}%
\usepackage{algorithmicx}%
\usepackage{algpseudocode}%
\usepackage{listings}%

\theoremstyle{thmstyleone}%
\theoremstyle{thmstyletwo}%

\theoremstyle{thmstylethree}%

\begin{document}

\title{Dynamic Behaviors of Materials Under Angle-Resolved Experiments: A Matrix Framework for Nonlinear Physics}

\author[1]{\fnm{Hongjin} \sur{Xiong}}\email{hongjin.xiong@connect.polyu.hk}
\author[1]{\fnm{Zhonghui} \sur{Xia}}\email{25114884r@connect.polyu.hk}
\author[2]{\fnm{Xuan} \sur{Ye}}\email{yexuan@cuhk.edu.cn}

\author*[1]{\fnm{Teng} \sur{Ma}}\email{tengma@polyu.edu.hk}

\affil[1]{\orgdiv{Department of Physics and Materials},\orgname{The Hong Kong Polytechnic University}, \orgaddress{ \city{Hong Kong}, \country{P.R. China}}}

\affil[2]{\orgdiv{School of Science and Engineering}, \orgname{The Chinese University of Hong Kong (Shenzhen)}, \orgaddress{\street{Longgang}, \city{Shenzhen}, \postcode{518172}, \state{Guangdong}, \country{P.R. China}}}

\abstract{Matter and external stimuli interact dynamically almost everywhere in nature. Under an applied field, a material seldom stays in the state in which it was characterized, and its properties change with time. Angle-resolved experiments are the standard way to resolve direction-dependent properties, however, current theoretical tools for data analysis are built for static states, leaving dynamic behaviors largely neglected. Here, we develop a generalized ``input-coefficient-output'' (ICO) approach that constructs spatial operation matrices and symmetry-reduced coefficient matrices across diverse physical systems, as we verify for nonlinear optics, elasticity, electrostriction and magnetostriction, and a Python program is released for convenient application. Using angle-resolved polarized second-harmonic generation (SHG) as an example, we focus on the dynamic modeling of material properties under angle-dependent stimuli, construct dynamic matrix equations that capture the time-dependent evolution of material responses, and propose three dynamic models and a point-to-point data fitting method. For the SHG experiments performed on $T_d-$WTe$_2$, the discrete dynamic model captures the asymmetric characteristics of the measured polar pattern with a significantly improved coefficient of determination ($R^2$ increasing from 0.567 to 0.773), whereas conventional static models are constrained to a symmetric curve. The evolution operator extracted from the dynamic model connects the macroscopic measurement to microscopic intermediate states during their interaction with stimuli. This work provides a practical matrix framework for dynamic modeling of material properties, paving the way for exploring complex nonlinear physics under realistic experimental conditions.}


\maketitle

\section{Introduction}

The dynamic interaction between matter and external stimuli is common in nature \cite{dynamic1, dynamic2, dynamic3,dynamic4, decay_irridation}. In real-world conditions, materials rarely remain static under external fields, causing their properties to change over time \cite{dynamic4,time1,time2,decay_irridation}. Therefore, capturing these dynamic processes is crucial for both fundamental research and practical applications of materials. 

To investigate direction-dependent properties of materials, angle-resolved measurements are ideal for probing linear and nonlinear phenomena \cite{angel_hall,angle-nonlinear_optical,angle_SHG,angle_SHG2,angle_THG,angle_magnetostriction,angle_raman}. By applying in-plane stimuli at various angles, the directional dependence of material properties can be effectively characterized. For example, polarized second-harmonic generation (SHG) is a common technique for studying second-order optical properties of materials \cite{angle-nonlinear_optical,angle_SHG,angle_SHG2,angle_THG}. Extracting physical properties from test data, however, requires an angular- dependent model, and deriving that model is not trivial, because handling rotation matrices under the Voigt notation \cite{Nonlinear_optics} (Eq.~\eqref{notation}) is inherently cumbersome. Consequently, some studies resort to specific computational methods. For instance, in the supplementary materials of Ref. \cite{angle-nonlinear_optical}, to calculate the vector $\vec{E}\vec{E}=(E_xE_x,E_yE_y,E_zE_z,2E_yE_z,2E_xE_z,2E_xE_y)^T$ under rotation, the authors employ a three-dimensional rotation matrix to obtain the rotated $\vec{E}=(E_x,E_y,E_z)^T$, and then manually construct $\vec{E}\vec{E}$ by multiplying each pair $E_iE_j~(i,j=x,y,z)$ individually. This procedure is not only computationally tedious but also hampers the study of dynamic material properties, as it requires differentiating each element of $\vec{E}\vec{E}$ individually when constructing dynamic equations. Moreover, previous works {\color{blue}may} have largely neglected dynamic behaviors of materials \cite{angle-nonlinear_optical,SHG_PNAS1,SHG_PNAS2}, relying instead on time-independent functions to extract only static properties. Therefore, it is essential to incorporate time-dependent characteristics under realistic experimental conditions.

In practical applications, materials are exposed to diverse scenarios, each associated with different spatial operation matrices (such as rotation matrix). Thus, the most effective way to derive dynamic functions is to construct matrix equations. However, before solving these dynamic problems, another obstacle is how to derive the high-order coefficient matrices for materials with specific symmetries and how to further operate on them.

In solid-state physics, the properties of matter are defined by the relationship between external stimuli and material responses, which are strictly governed by crystal symmetry \cite{Symmetry_Strategy,Generalized_Neumann_Principle,Symmetry-related_properties_of_plasmon-polaritons,Tunneling_properties_T3}. These stimulus-response relations are generally represented by coefficient matrices \cite{Physical_properties_of_crystals,Nonlinear_optics,newnham,International_tables_for_crystallography,Property_tensors}, and analyzing how symmetry operations constrain their elements is essential for probing material properties. For linear phenomena, the coefficient matrix is square and can be transformed straightforwardly via a similarity transformation. In contrast, for higher-order nonlinear phenomena, such as nonlinear optics \cite{Nonlinear_optics}, the nonlinear Hall effect \cite{Nonlinear_hall_effects}, and magnetostriction \cite{newnham,magnetostriction,Magnetostriction_2}, the dimensions of stimuli and responses differ, making the coefficient matrices rectangular or excessively large and difficult to transform through simple similarity operations.

The standard route to derive those matrices is the ``direct inspection method'' \cite{direct-inspection_method1,direct-inspection_method2}, which is prone to errors for complex symmetries or higher-order nonlinear phenomena. For instance, Table 15.5 of Ref. \cite{newnham} lists ``$N_{45}=N_{54}$'' for the magnetostriction tensor of $C_3$ crystals, contradicting Eq. (92) of Ref. \cite{magnetostriction}, where $N_{45}=-N_{54}$. Similarly, Table 8 of Ref. \cite{Physical_properties_of_crystals} and Table 12.2 of Ref. \cite{newnham} report ``$d_{16}=-2d_{22}$'' for $C_3$ and $3m$ crystals, whereas Eq. (A5) of Ref. \cite{C3_piezoelectric_tensor_violate} and Fig. 2 of Ref. \cite{3m_piezoelectric_tensor_violate} give ``$d_{16}=-d_{22}$''. To overcome these limitations, tensor-based methods have been proposed \cite{Walpole,magnetostriction,C3_piezoelectric_tensor_violate,eigenvector_approach_for_deriving_non-vanishing_tensor_elements}. but each method relies on a different highly specialized mathematical technique, limiting accessibility for experimental scientists. Also, each work focuses on only a single material property, making generalization difficult, and they provide only static coefficient matrices without corresponding operation matrices, which hinders the construction of dynamic matrix equations for materials under realistic conditions.

Therefore, to capture the dynamic behavior of materials under various stimuli, enable scalable research into higher-order nonlinear physics, and facilitate experimental work, we first develop an ``input-coefficient-output'' (ICO) approach for constructing spatial operation matrices for material property coefficient matrices. By applying this approach, the reduced coefficient matrices of certain symmetric materials across various physical systems can be derived. The ICO approach provides a concise notation that allows researchers to intuitively reason about spatial operations. It requires only the orders of the input and output vectors, and a Python program (Section \ref{ICO_chi}) is provided for convenient application. We check the ICO formalism (Section \ref{3m_symmetry} of SM) by deriving reduced susceptibility matrices of representative crystal systems (Section \ref{sec:verification} of SM), showing excellent agreement with established results, and extend the ICO formalism to elasticity, electrostriction, and magnetostriction (Section \ref{Other physical systems} of SM). Based on this formalism, we then investigate the dynamic properties of materials. We consider three angle-resolved polarized SHG experiment models (Section \ref{dynamic problem}), build the dynamic matrix equations, and derive analytical solutions. Unlike conventional curve fitting, which enforces a symmetric profile and fails to capture the asymmetric experimental data, our point-to-point method captures this asymmetry in the measured polar pattern (Section \ref{SHG_test}), which is further confirmed by the coefficient of determination ($R^2$). Furthermore, the evolution operator extracted from our dynamic model provides a bridge between macroscopic phenomenological observations and microscopic mechanisms. By combining it with structural characterization and first-principles calculations, this framework can be extended to locate the intermediate states that a material passes through under external stimuli. We propose a conceptual workflow to illustrate this promising direction. 

Therefore, this work addresses the dynamic modeling of material properties under various stimuli. We establish a generalized ICO framework to efficiently construct and manipulate high-order coefficient matrices, and further propose three dynamic models and a point-to-point fitting strategy for experimental data. Furthermore, the evolution operator extracted from our dynamic model provides a macroscopic bridge to microscopic mechanisms. By bridging clear physical insights with straightforward mathematics, this work paves the way for solving complex dynamic problems and provides a versatile tool for exploring symmetry-constrained phenomena in realistic experimental conditions \cite{angle-nonlinear_optical,angle_SHG,angle_SHG2,angle_THG,angle_magnetostriction}.

\section{General ICO formalism }\label{ICO_chi}

To realize the construction of dynamic model, we first propose the ICO formalism to manipulate the higher-order static properties of materials. 

Consider three-dimensional external stimulus and response vectors, $\vec{E}_n$ and $\vec{J}$, which are connected by a coefficient matrix $\chi^{(1\times n)}$
\begin{equation}
    \vec{J} = \chi^{(1\times n)} \cdot \vec{E}_1 \vec{E}_2 \cdots \vec{E}_n,
    \label{i_order_function}
\end{equation}
where $\vec{J}\equiv(J_x,J_y,J_z)^T$ and $\vec{E}=(E_x,E_y,E_z)^T$. The superscript $(1\times n)$ indicates that the coefficient matrix connects an $n$-th order stimulus to a first-order response. Eq.~\eqref{i_order_function} is commonly used in nonlinear optics \cite{Nonlinear_optics}. In component form, it is written as
$J_l =\sum_{ijk}\chi_{lijk} E_{1i} E_{2j}...E_{n k}$, where $l,i,j,k,\ldots=x,y,z$. To describe higher-order responses and construct the coefficient matrix $\chi$ more easily, we rewrite this relation using the Kronecker product:
\begin{equation}
    \vec{J}^{\otimes m} = \chi^{(m\times n)} \cdot \vec{E}^{\otimes n},
    \label{generalized_high_order_function}
\end{equation}
where the response is generalized from first order to $m$-th order, and the coefficient matrix is changed accordingly. Here,
\begin{align}
  \vec{E}^{\otimes n}
  &\equiv
\vec{E}_1 \otimes \vec{E}_2 \cdots \otimes \vec{E}_n
  =
\vec{E}_1 \vec{E}_2 \cdots \vec{E}_n,\label{def1}
\\
  \vec{J}^{\otimes m}
  &\equiv
  \vec{J}_1 \otimes \vec{J}_2 \cdots \otimes \vec{J}_m.\label{def2}
\end{align}
Eq.~\eqref{generalized_high_order_function} allows us to describe more complicated systems with the $m$-th order response, such as the elastic system discussed in Sec.~\ref{Other physical systems}.

\begin{figure}[htbp]
    \centering
 \includegraphics[width=1\linewidth]{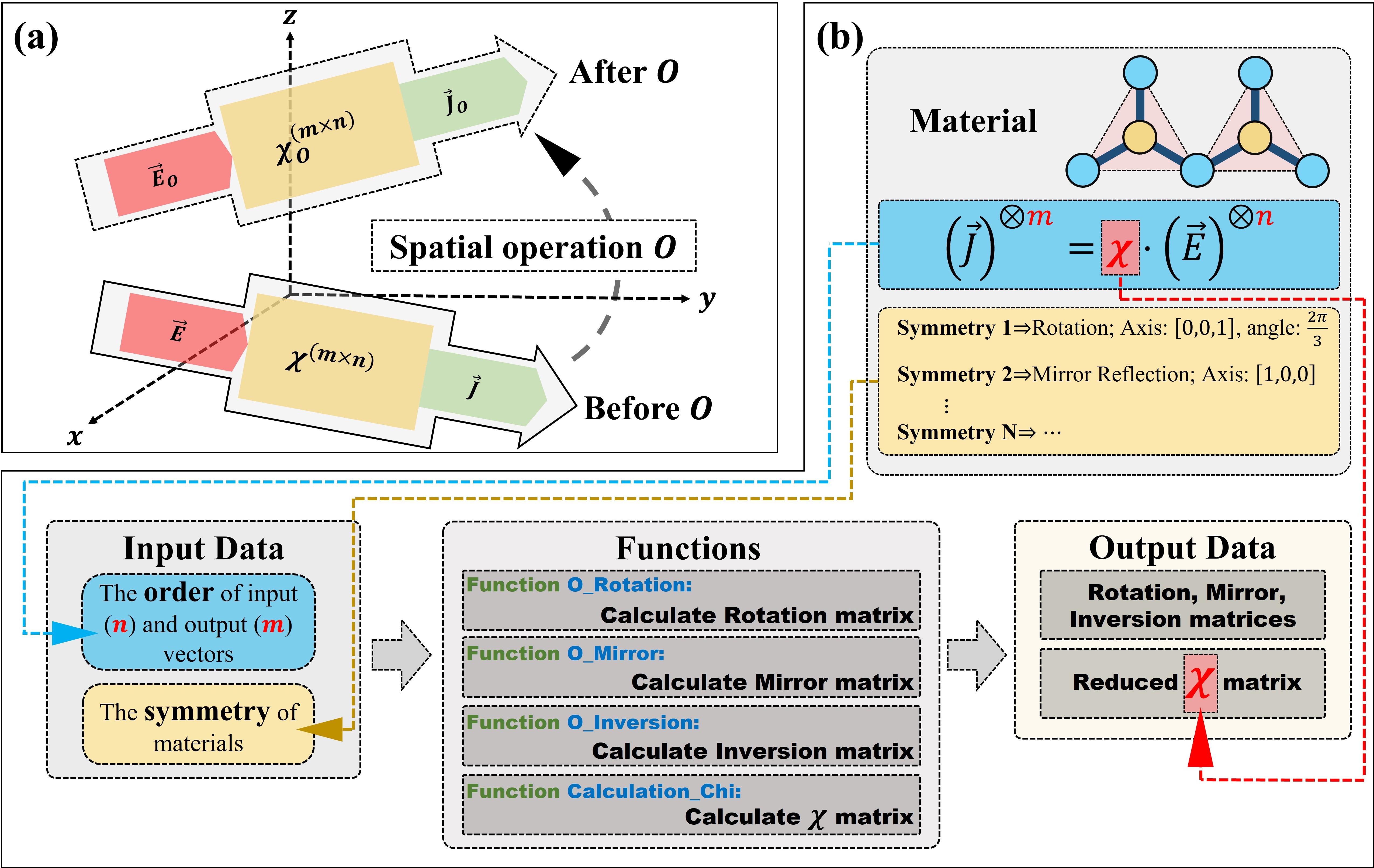}
    \caption{(a) Diagram of the Input-Coefficient-Output formalism showing invariance under spatial transformations $O$. (b) The workflow of the ``$\chi$ program''.}
    \label{fig:combined_figure}
\end{figure}

In the following, based on Eq.~\eqref{generalized_high_order_function}, we derive the transformation law for the coefficient matrix under a spatial operation $O$ such as rotation, inversion, or mirror reflection. As illustrated in Fig.~\ref{fig:combined_figure}a, the ICO formalism in Eq.~\eqref{generalized_high_order_function} should remain invariant under spatial operations. Therefore, after a spatial operation, Eq.~\eqref{generalized_high_order_function} remains valid in the form:
\begin{equation}
    (\vec{J}_O)^{\otimes m} =\chi^{(m\times n)}_O \cdot (\vec{E}_O)^{\otimes n}, \label{high_order_relation_operation}
\end{equation}
where $\vec{J}_O \equiv O \cdot \vec{J}$ and $\vec{E}_O \equiv O \cdot \vec{E}$, and the coefficient matrix after symmetric operation is denoted as $\chi^{(m\times n)}_O$, which will be determined in what follows.

By applying the property $(A \cdot \vec{u}) \otimes (B \cdot\vec{v}) = (A \otimes B)\cdot(\vec{u} \otimes \vec{v})$ and the definitions in Eq.~\eqref{def1} and Eq.~\eqref{def2}, Eq.~\eqref{high_order_relation_operation} can be rewritten as
\begin{equation}
   O^{\otimes m} \cdot \vec{J}^{\otimes m} =\chi^{(m\times n)}_O \cdot O^{\otimes n} \cdot \vec{E}^{\otimes n}, \label{high_order_relation_operation_expanded}
\end{equation}
with $O^{\otimes m}$ defined as $O_1 \otimes O_2 \cdots \otimes O_m$. Comparing this result with Eq.~\eqref{generalized_high_order_function}, we find that $\chi^{(m\times n)} = (O^{-1})^{\otimes m} \cdot \chi^{(m\times n)}_O \cdot O^{\otimes n}$, which leads to
\begin{equation}
    \chi^{(m\times n)}_O = O^{\otimes m} \cdot \chi^{(m\times n)} \cdot (O^{-1})^{\otimes n}, \label{O_chi}
\end{equation}
where $O^{-1}$ is the inverse operator of $O$.
Equation \eqref{O_chi} shows how the coefficient matrix, which connects an 
n-th order stimulus and an m-th order response, changes under the spatial operation 
$O$. For a material with a specific symmetry that is invariant under operation 
$O$, we have:
\begin{equation}
    \chi^{(m\times n)} - O^{\otimes m} \cdot \chi^{(m\times n)} \cdot (O^{-1})^{\otimes n}=0. \label{dede}
\end{equation}

For convenience, we define the widely used $m$-th order rotation, inversion, and mirror reflection operators as $[R_i(\phi_i)]^{\otimes m}$, $I^{\otimes m}$, and 
$M_i^{\otimes m}$, respectively. The specific definitions of these $m$-th order operators are provided in Section~\ref{construct} of the SM.
Note that the subscript $i$ for the rotation operator takes values in $x, y, z$, representing a rotation by an angle $\phi_i$ about the $i$-axis. The subscript $i$ for the mirror reflection takes values in $xy, yz, zx$, representing a mirror reflection about the $i$-plane.

Given the $m$-th order operator, one can solve Eq.~\eqref{dede} to find the coefficient matrix for a certain symmetric material. Since the equation becomes a large-scale matrix equation when the order is large, we provide a Python program called the ``$\chi$ program'' to solve it \cite{chipackage}. The algorithm for this program is shown in Fig.~\ref{fig:combined_figure}b.

This code requires two inputs. The first input is the order of the stimulus $(n)$ and response $(m)$. The second input is the symmetry of the material, specifically the spatial operators $O$ that leave the material invariant.

When the program runs, it first calls the  ``Calculation$_-$Chi'' function (Functions box, Fig.~\ref{fig:combined_figure}b). This function calls other functions (``O$_-$Rotation'', ``O$_-$Mirror'' and ``O$_-$Inversion'', Fig.~\ref{fig:combined_figure}b) to create spatial operation matrices based on the input data. Then, the ``Calculation$_-$Chi'' function solves Eq. \eqref{dede}. Finally, 
the ``$\chi$ program'' provides two outputs. The first is the matrix form of $O^{\otimes m}$ and  $(O^{-1})^{\otimes n}$, and the second is the solution for the symmetry-constrained coefficient matrix $\chi$. 
The coefficient matrices $\chi$ used in this paper are obtained by running this program. 

To verify the ICO formalism, we calculate the coefficient matrices for both second-order stimuli systems (with first- and second-order responses) and various representative crystal systems (including triclinic, monoclinic, orthorhombic, and tetragonal crystals). We demonstrate that the derived reduced matrices perfectly reproduce the established results in the literature. Furthermore, we use the 
``$\chi$ program" to resolve several discrepancies reported in prior studies. Comprehensive derivations and validations are provided in Section \ref{Validation_sm} of SM.

\section{Dynamic Problem}\label{dynamic problem}
\begin{figure}[h]
    \centering
    \includegraphics[width=1\linewidth]{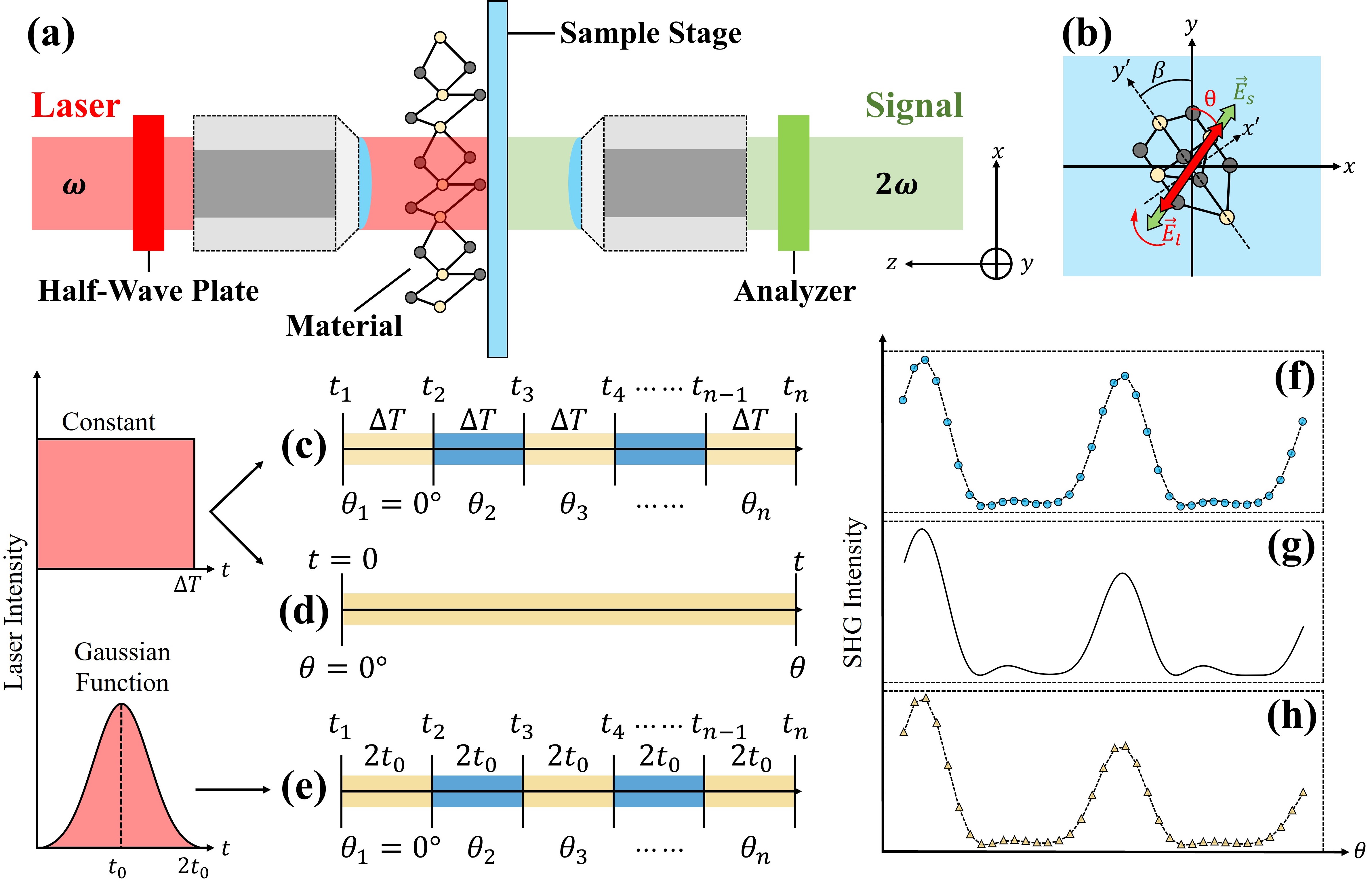}
    \caption{Schematic of SHG experiment. (a) Optical path diagram. (b) {Side view of Fig. a}. The red and green arrows ($\vec{E}_l\parallel\vec{E}_s$) represent the vectors of laser and signal analyzer, respectively.  The intensity of laser is a constant and: (c) $\vec{E}_l$ and $\vec{E}_s$ stay at each angle $\theta_n$ for $\Delta T$ seconds and then turn to the next angle $\theta_{n+1}$ immediately; (d) $\vec{E}_l$ and $\vec{E}_s$ rotating continuously. (e) The intensity of laser is a Gaussian function and $\vec{E}_l$ and $\vec{E}_s$ stay at each angle $\theta_n$ for $2t_0$ seconds and then turn to the next angle $\theta_{n+1}$ immediately. (f)-(h) the SHG intensity as a function of $\theta$ for the scenarios shown in (c)-(e), respectively.}
    \label{fig:dynamic}
\end{figure}

The ICO formalism enables the straightforward construction of operation matrices for higher-order properties of materials, thereby indicating that the dynamic behavior of matter under various operating conditions can be readily modeled. To illustrate this, we take the angle-resolved polarized SHG experiment as an example. As shown in Fig.~\ref{fig:dynamic}a, the material receives the stimuli of a polarized-laser and then emits a signal, these are denoted by the red and green area, respectively. To analyze the properties of materials, we derive the equation of the intensity of signal, $I_{\text{SHG}}$. In this section, we propose three test conditions and establish the corresponding models for $I_{\text{SHG}}$ with the aid of the ICO formalism. Then, we apply the model developed in Section \ref{Discrete Model} to analyze our SHG experimental data (Section \ref{SHG_test}).  

In Fig.~\ref{fig:dynamic}b, assume that there is an angle $\beta$ between the axis of material and lab coordinate ($y'$ and $y$), and the vector $\vec{E}_l$ and $\vec{E}_s$ represent the polarized laser and signal analyzer, respectively. The $\vec{E}_l$ always parallel to $\vec{E}_s$ and rotates clockwise from $\theta=0^\circ$. Based on the $y'$ axis of materials (set $y'=(0,1)^T$), the $\vec{E}_l$ and $\vec{E}_s$ are (both $\vec{E}_l$ and $\vec{E}_s$ haven't $z$ component):
\begin{equation}
\vec{E}_{l}=\vec{E}_{s}=R_z^T(\theta+\beta)\cdot\vec{y}~'=\begin{pmatrix}
        \cos(\theta+\beta)&\sin (\theta+\beta)\\  
        -\sin(\theta+\beta)&\cos(\theta+\beta)
    \end{pmatrix}\cdot\begin{pmatrix}
        0\\  
        1
    \end{pmatrix}=\begin{pmatrix}
        \sin(\theta+\beta)\\  
        \cos(\theta+\beta)
    \end{pmatrix}.\label{ElEs}
\end{equation}

The scalar polarization of the SHG signal (the ``Signal'' of Fig.~\ref{fig:dynamic}a) is given by:
\begin{align}
    P_{\text{lab}}&=\vec{E}_s^T\cdot \vec{P}_{\text{m}}\equiv\vec{E}_s^T\cdot[\chi^{(2)}\cdot\vec{E}_{l}\vec{E}_{l}]= [R_z(\theta+\beta)^T\cdot{\vec{y}~'}]^T \cdot\bigg[\chi^{(2)}\cdot [R_z^T(\theta+\beta)]^{\otimes 2}\cdot\big(\vec{y}~'\otimes\vec{y}~'\big)\bigg],\label{polarized_SHG}
\end{align}
where $\vec{P}_{\text{m}}$ is the polarization vector emitted by materials. The $P_{\text{lab}}$ (scalar) is the $\vec{P}_{\text{m}}$ after processing by the analyzer (green box, Fig.~\ref{fig:dynamic}a). By calculating $I_{\text{SHG}}=\left|P_{\text{lab}}\right|^2$, we obtain the intensity function of the SHG experiment. This is how the conventional model is built in the previous study \cite{angle-nonlinear_optical}.

To model the dynamic behavior of materials, we assume the material properties variant under laser exposure, which characterized by $\chi^{(2)}$. Thus, consider the equation:
\begin{align}
    \frac{d{\chi^{(2)}}}{dt}&=\big[a\sin^2(\theta+\beta)+b\cos^2(\theta+\beta)\big]\cdot\big(\vec{E}_l\cdot \vec{E}_l^T\big)\cdot{\chi^{(2)}}=A(\theta+\beta)\cdot{\chi^{(2)}},\label{decay_chi}
\end{align}
where $a$ and $b$ represent the evolution rates along the $\vec{y}~'$ and $\vec{x}~'$ directions, respectively. For simplicity, we assume them to be constant, although they should in general be functions.

Using this equation to illustrate the evolution of materials under irradiation is reasonable because the eigenvectors and eigenvalues of $A(\theta+\beta)$ are:
\begin{align}
    \vec{v}_{\parallel}=\vec{E}_l=\begin{pmatrix}
         \sin(\beta+\theta)\\  
        \cos(\beta+\theta)
    \end{pmatrix},~ \lambda_{\parallel}=a\sin^2(\theta+\beta)+b\cos^2(\theta+\beta);~\vec{v}_{\bot}=\begin{pmatrix}
         -\cos(\beta+\theta)\\
         \sin(\beta+\theta)
    \end{pmatrix},~ \lambda_{\bot}=0.
\end{align}
Given that $\vec{v}_{\parallel}=\vec{E}_l$, $\vec{E}_l\bot \vec{v}_{\bot}$, $\lambda_\parallel\neq0$ and $\lambda _\bot=0$, the laser exerts an effect exclusively along its polarization direction ($\vec{E}_l$). If $a\neq b$, the $\lambda_\parallel$ represents the evolution rate of anisotropic materials. When $\vec{E}_l$ is aligns with the $x'$ or $y'$ axis of the material, the $\lambda_\parallel$ takes the values $a$ or $b$, respectively. In the case where $a=b$, we have $\lambda_\parallel=a$, which illustrates the isotropic behavior of materials.

At any $\theta_n$, when $A(\theta+\beta)$ is time-independent, the solution of Eq.~\eqref{decay_chi} is:
\begin{align}
\chi^{(2)}(t)=e^{A(\theta_n+\beta)t}\cdot\chi^{(2)}_{i},
\end{align}
where the $\chi^{(2)}_{i}$ is the $\chi^{(2)}$ at initial state, and the $e^{A(\theta_n+\beta)t}$ is (refer to Eq.~\eqref{decay_chi}):
\begin{equation}
    e^{A(\theta_n+\beta)t}=\mathbb{I}_{2\times 2}+ [e^{\lambda_\parallel~t }-1]\cdot\big(\vec{E}_l\cdot \vec{E}_l^T\big). \label{eAt}
\end{equation}

Therefore, in Section \ref{Discrete Model}, by substituting Eq.~\eqref{eAt} into Eq.~\eqref{polarized_SHG} to calculate $P_{\text{lab}}$ and evaluating its squared magnitude $|P_{\text{lab}}|^2$, the SHG intensity $I_{\text{SHG}}$ can be obtained. However, in Sections \ref{Continue model} and \ref{dis_gaussian_main}, $t$ is a variable in $A(\theta+\beta)$, so Eq.~\eqref{eAt} cannot be used. Instead, we calculate the expression for $\frac{d\vec{P}_{\text{m}}}{dt}$, substitute Eq.~\eqref{decay_chi} into it, solve the resulting differential equations, and evaluate $|P_{\text{lab}}|^2$ to obtain $I_{\text{SHG}}$.  

\subsection{Discrete model}\label{Discrete Model}
Generally, the angle-resolved polarized SHG experiment measures materials under intermittent conditions (refer to Fig.~\ref{fig:dynamic}c, the intensity of laser is a constant). First, at polarization angle $\theta_1=0^\circ$, the material is stimulated by laser for $\Delta T$ seconds, then, the polarization angle is changed to $\theta_2$, and the laser is performed again for $\Delta T$ seconds, and so on. At each $\theta_n$, according to Eq.~\eqref{polarized_SHG}, the $\vec{P}_{\text{m}}$ is:
\begin{equation}
    \vec{P}_{\text{m}}(t)=\chi^{(2)}(t)\cdot \big[R_z^T(\theta_n+\beta)\big]^{\otimes 2}\cdot(\vec{y}~')^{\otimes 2}.\label{Pm_discrete}
\end{equation}
Thus, the process is, first, when $\theta=\theta_1=0^\circ$, initial $\chi^{(2)}=\chi^{(2)}_{t=0}$, during the time of $t_0\leq t\leq t_1 $. According to the Eq.~\eqref{eAt}, we have:
\begin{align}
\chi^{(2)}(t)=e^{A(\theta_1+\beta)t}\cdot\chi^{(2)}_{t=0}.\label{dchi_0}
\end{align}
Thus, refer to Eq.~\eqref{Pm_discrete}, the $\vec{P}_{\text{m}}$ is: 
\begin{align}
\vec{P}_{\text{m}}=e^{A(\theta_1+\beta)t}\cdot\chi^{(2)}_{t=0}\cdot \big[R_z^T(\theta_1+\beta)\big]^{\otimes 2}\cdot(\vec{y}~')^{\otimes 2}.
\end{align}
When $t=\Delta T$, the system changes to the next step ($\theta=\theta_2$), and the initial $\chi^{(2)}=\chi^{(2)}_{t=\Delta T}$, during the time of $t_1\leq t\leq t_2$, we have:
\begin{align}
\chi^{(2)}(t)&=e^{A(\theta_2+\beta)t}\cdot\chi^{(2)}_{t=\Delta T}=e^{A(\theta_2+\beta)t}\cdot e^{A(\theta_1+\beta)\Delta T}\cdot\chi^{(2)}_{t=0},
\end{align}
and:
\begin{align}
\vec{P}_{\text{m}}&=e^{A(\theta_2+\beta)t}\cdot e^{A(\theta_1+\beta)\Delta T}\cdot\chi^{(2)}_{t=0}\cdot \big[R_z^T(\theta_2+\beta)\big]^{\otimes 2}\cdot(\vec{y}~')^{\otimes 2}.
\end{align}

Thus, for the step $n$:
\begin{align}
    \vec{P}_{\text{m}}=e^{A(\theta_{n}+\beta)t}\cdot \left[\prod_{i=0}^{n-1}  e^{A(\theta_{i}+\beta)\Delta T}\right]\cdot\chi^{(2)}_{t=0}\cdot \big[R_z^T(\theta_n+\beta)\big]^{\otimes 2}\cdot(\vec{y}~')^{\otimes 2}.
\end{align}
If using the average value to represent the test results for polarization angle ($\theta=\theta_n$):
\begin{align}
    \overline{\vec{P}_{\text{m}}}=\bigg[\frac{1}{\Delta T}\int_0^{\Delta T} e^{A(\theta_{n}+\beta)t}~ dt\bigg]\cdot\left[\prod_{i=0}^{n-1}  e^{A(\theta_{i}+\beta)\Delta T}\right]\cdot\chi^{(2)}_{t=0}\cdot \big[R_z^T(\theta_n+\beta)\big]^{\otimes 2}\cdot(\vec{y}~')^{\otimes 2},
\end{align}
with:
\begin{align}
    \frac{1}{\Delta T}\int_0^{\Delta T} e^{A(\theta_n+\beta)t}~ dt&= \mathbb{I}_{2\times 2}+ \bigg(\frac{1}{\Delta T}\cdot\frac{e^{\lambda_\parallel\Delta T}-1}{\lambda_\parallel}-1\bigg)\cdot\big(\vec{E}_l\cdot \vec{E}_l^T\big) .
\end{align}

Therefore, refer to Eq.\eqref{polarized_SHG}, the SHG intensity $I_{\text{SHG}}$ at $\theta_n$ is:
\begin{equation}
    I_{\text{SHG}}(\theta_n)=\left|P_{\text{lab}}(\theta_n)\right|^2=\left|\vec{E}_l(\theta_n)^T\cdot \overline{\vec{P}_{\text{m}}}(\theta_n)\right|^2.\label{discrete_intensity}
\end{equation}
Its scatters are shown in Fig.~\ref{fig:dynamic}f. 

\subsection{Continue model}\label{Continue model}
We further assume the materials working under continue condition, suppose that the polarized laser rotating with angular velocity is $\omega=\frac{d\theta}{dt}$ (Fig.~\ref{fig:dynamic}d), and $\chi^{(2)}(t)|_{t=0}=\chi^{(2)}_0$. According to Eq.~(\ref{polarized_SHG}) and Eq.~(\ref{eAt}), we have:
\begin{align}
    \vec{P}_{\text{m}}(t)&= \chi^{(2)}(t)\cdot \big[R_z^T\big(\theta(t)+\beta\big)\big]^{\otimes 2}\cdot(\vec{y}~')^{\otimes 2}.\label{Pc_continue}
\end{align}
To illustrate the relation between $I_{\text{SHG}}$ and $\theta$, differentiating both sides of Eq.~\eqref{Pc_continue} with respect to $t$, and applying $\frac{d\theta}{dt}=\omega$, we have:
\begin{align}
    \frac{d\vec{P}_\text{m}}{d\theta}= B(\theta)\cdot \vec{P}_\text{m}+l(\theta),\label{dpc_continue}
\end{align}
with:
\begin{align}
     B(\theta)\equiv \frac{1}{\omega}\cdot A(\theta+\beta),~~
         l(\theta)\equiv\chi^{(2)}\cdot\frac{d~\big[ R_z^T(\theta+\beta)\big]^{\otimes 2}}{d\theta}\cdot\big(\vec{y}~'\otimes\vec{y}~'\big).
\end{align}
Eq.~\eqref{dpc_continue} is equivalent to a continuous-time linear time-varying (CLTV) system \cite{Linear_systems_theory}, and the solution is:
\begin{align}
    \vec{P}_{\text{m}}=\Phi(\theta,0)\cdot\vec{P}_m(0)+\int_{0}^\theta \Phi (\theta,\alpha)\cdot l(\alpha)\cdot d\alpha,
\end{align}
where, $\Phi(\theta,0)$ is Peano-Baker series \cite{Linear_systems_theory}.

Finally, the SHG intensity can be obtained by calculating: 
\begin{equation}
I_{\text{SHG}}=\left|\vec{P}_{\text{lab}}\right|^2=\left|\vec{E}_s^T[\theta(t)]\cdot\vec{P}_{\text{m}}[\theta(t)]\right|^2
\end{equation}
And the correspondent picture is Fig.~\ref{fig:dynamic}g. 

\subsection{Discrete model with time dependent stimuli}\label{dis_gaussian_main}
Based on the discrete model, suppose that if the laser is a Gaussian pulse (Fig. \ref{fig:dynamic}e), the mathematical expression of its intensity is \cite{Gaussian_pulse}:
\begin{equation}
    a(t)=a_0~e^{-\frac{(t-t_0)^2}{\tau^2}}.
\end{equation}
Here, $a_0$ is the peak amplitude, $t_0$ is the center position, $\tau$ is the width parameter. And the $\vec{P}_{\text{m}}$ is:
\begin{align}
    \vec{P}_{\text{m}}&=a(0)^2\cdot[e^{-\frac{2}{\tau^2}(t^2-2t_0t)}\cdot e^{k(t)\cdot A(\theta_n+\beta)}]\cdot\left[ \prod_{i=0}^{n-1} e^{k(2t_0)\cdot A(\theta_i+\beta)}\right]\cdot\chi^{(2)}_{t=0}\cdot \big[R_z^T(\theta_n+\beta)\big]^{\otimes 2}\cdot(\vec{y}~')^{\otimes 2}.
\end{align}
the $I_{\text{SHG}}$ at $\theta_n$ is:
\begin{equation}
    I_{\text{SHG}}(\theta_n)=\left|\vec{E}_l(\theta_n)^T\cdot \vec{P}_{\text{m}}(\theta_n,t_0)\right|^2.
\end{equation}
Its scatters are shown in Fig.~\ref{fig:dynamic}h
(more details, see section \ref{dis_gaussian} of SM.).

\section{SHG Experiment} \label{SHG_test}
\begin{figure}[h]
    \centering
    \includegraphics[width=1\linewidth]{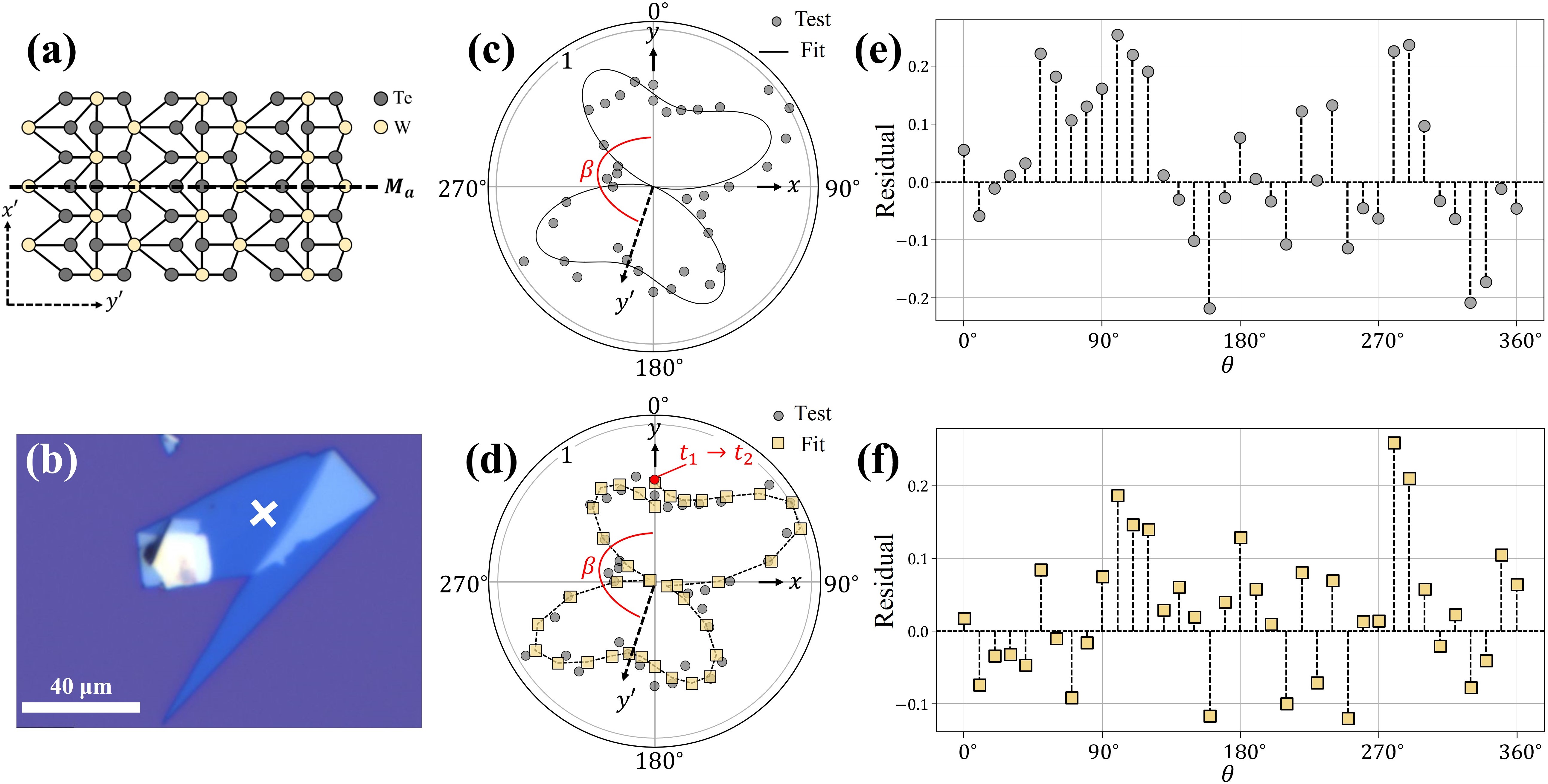}
    \caption{SHG experiment. (a) Crystal structure of $T_d-$WTe$_2$, the $M_a$ is a mirror plane, parallel to $yz$ plane. (b) Optical image of the $T_d-$WTe$_2$ sample, $\times$ is the SHG test point. (c) SHG test results (circular scatters) and fitting by conventional time independent model (Eq.~(\ref{conventioanl_model}), black solid line). (d) SHG test results (gray scatters) and fitting by the model of section \ref{Discrete Model} (yellow boxes). (e) and (f) The residuals of the fitting results in Fig. c and d, respectively. The $x$ and $y$ axes are the coordinate of lab, the $y'$ axis represents the direction of material. The angle $\beta$ between $y$ and $y'$ axes are $162.11^\circ$ in Fig. c and $159.65^\circ$ in Fig. d}
    \label{fig:SHG}
\end{figure}

To show that the ICO method-based dynamic equations enable researchers to extract material properties more accurately, we perform angle-resolved polarized SHG experiments on $T_d-$WTe$_2$ (a two-dimensional material, Fig.~\ref{fig:SHG}a and \ref{fig:SHG}b) and analyze the data with both conventional model and discrete model from section \ref{Discrete Model}.

According to Ref. \cite{Wte2_Nature}, $T_d-$WTe$_2$ has only a mirror plane perpendicular to its $x'$ axis (Fig.~\ref{fig:SHG}a). Thus, using the ``$\chi$ program'' and considering permutation symmetry ($\chi_{ijk}=\chi_{ikj}$ \cite{Nonlinear_optics}), we obtain the second-order susceptibility tensor of $T_d-$WTe$_2$:
\begin{equation}
    \chi^{(2)}=\left(\begin{matrix}0 & \chi_{xxy} & \chi_{xxz} & \chi_{xyx}=\chi_{xxy} & 0 & 0 & \chi_{xzx} & 0 & 0\\\chi_{yxx} & 0 & 0 & 0 & \chi_{yyy} & \chi_{yyz} & 0 & \chi_{yzy}=\chi_{yyz} & \chi_{yzz}\\\chi_{zxx} & 0 & 0 & 0 & \chi_{zyy} & \chi_{zyz} & 0 & \chi_{zzy}=\chi_{zyz} & \chi_{zzz}\end{matrix}\right),
\end{equation}
which aligns with the result of Ref. \cite{Wte2_Prb}. As discussed in previous sections, since the $\vec{P}_{\text{m}} $ and $\vec{E}_l$ haven't the $z$ component, the $\chi^{(2)}$ can be reduced to:
\begin{equation}
    \chi^{(2)}=\left(\begin{matrix}0 & \chi_{xxy}  & \chi_{xxy} & 0 \\\chi_{yxx} & 0  & 0 & \chi_{yyy} \end{matrix}\right).
\end{equation}

To create the fitting equation of SHG test. Applying ICO formalism and refer to Ref. \cite{Wte2_Prb}, we first consider the conventional time independent model:
\begin{align}
    I_{\text{in}}(\theta)&=\left| \vec{E}_s^T \cdot \chi^{(2)} \cdot \big[R_z^T(\theta+\beta)\big]^{\otimes 2} \cdot \vec{E}_l\vec{E}_l \right|^2=\left|(2 \chi_{xxy}+\chi_{yxx} )\sin^{2}{\left(\beta + \theta \right)} \cos{\left(\beta + \theta \right)} + \chi_{yyy} \cos^{3}{\left(\beta + \theta \right)}\right|^2,\label{conventioanl_model}
\end{align}
where, the $\vec{E}_s$ and $\vec{E}_l$ are defined by Eq.~\eqref{ElEs}. And the discrete model is $I_{\text{dis}}(\theta_n)$ (Eq.~\eqref{discrete_intensity}). 

Subsequently, we normalized the experimental data (circular markers in Figs.~\ref{fig:SHG}c and \ref{fig:SHG}d) and fitted them using $I_{\text{in}}(\theta)$ (black solid lines in Fig.~\ref{fig:SHG}c) and $I_{\text{dis}}(\theta_n)$ (yellow boxes in Fig.~\ref{fig:SHG}d). As shown in Figs.~\ref{fig:SHG}e and \ref{fig:SHG}f, the average absolute residuals of the fits in Figs.~\ref{fig:SHG}c and \ref{fig:SHG}d are $0.103$ and $0.079$, respectively. The coefficients of determination ($R^2$) for Figs.~\ref{fig:SHG}c and \ref{fig:SHG}d are $0.567$ and $0.773$, respectively. These results indicate that the discrete model describes the material properties better than the conventional model.

This improvement arises because material properties evolve during the scan, causing the SHG intensity to drift and the measured polar pattern to become asymmetric. This phenomenon is commonly observed in experiments \cite{SHG_PNAS1,SHG_PNAS2}. The conventional model, constrained to a symmetric profile in polar coordinates, cannot capture this drift. In contrast, the discrete model can follow the evolution because it adopts a point-by-point fitting strategy and incorporates dynamic parameters ``$a$'' and ``$b$'' (see Eq.~\eqref{decay_chi}). By incorporating the dynamic properties of the material, this model provides a more practical mathematical expression for each data point. 

\begin{figure}[h]
    \centering
    \includegraphics[width=1\linewidth]{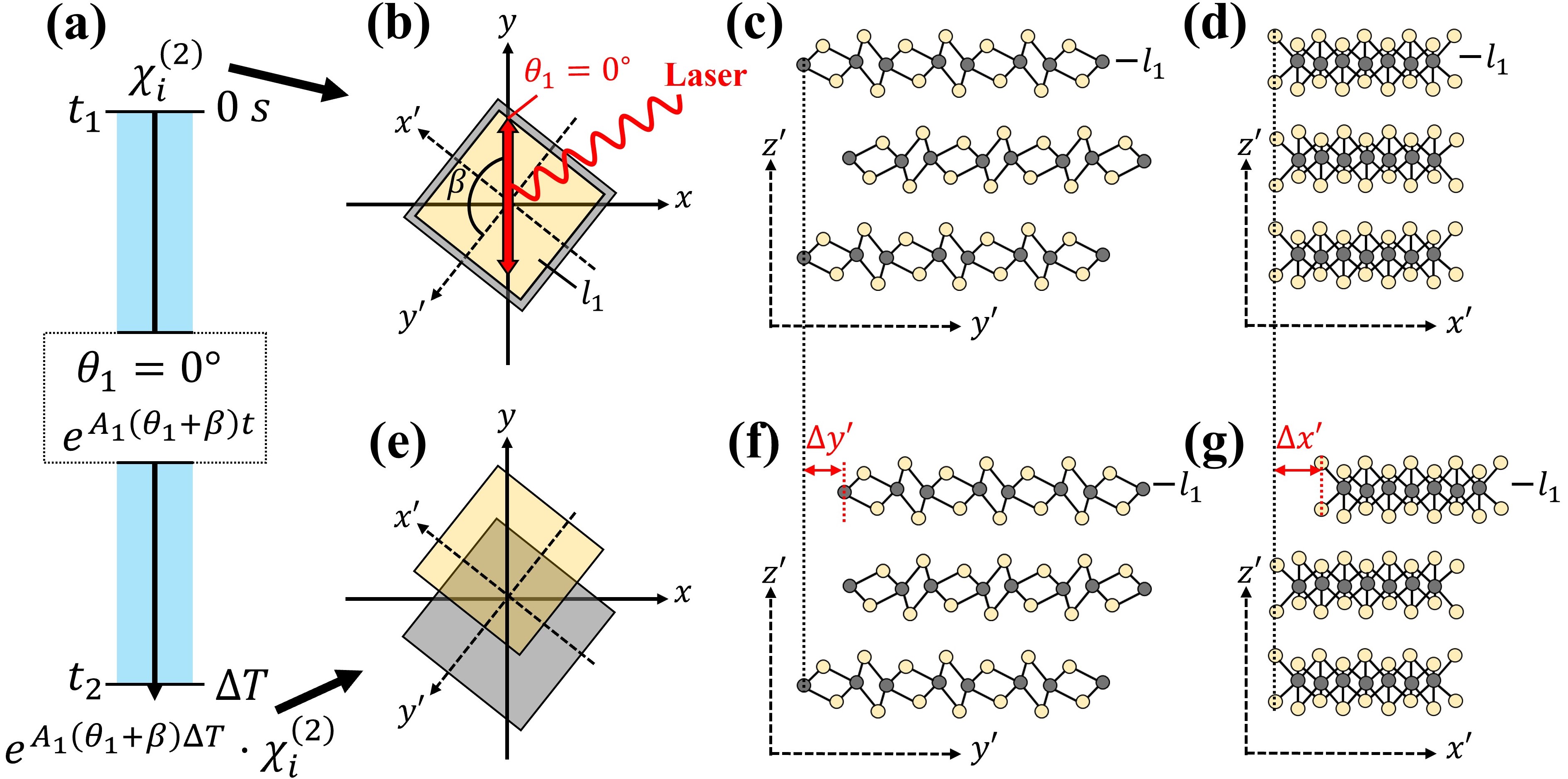}
    \caption{Conceptual illustration of the dynamic evolution and intermediate state investigation under external stimuli. (a) Timeline of the process at the first measurement point (red point in Fig.~\ref{fig:SHG}d), where the initial susceptibility tensor $\chi^{(2)}_i$ evolves into $e^{A_1(\theta_1+\beta)\Delta T} \cdot \chi^{(2)}_i$ over a time interval $\Delta T$. (b) The unperturbed $T_d-$WTe$_2$ at $t_1=0$ s with a susceptibility tensor of $\chi^{(2)}_i$. The dashed axes ($x'$ and $y'$) represent the material coordinate system, and the solid axes ($x$ and $y$) denote the laboratory coordinate system. The yellow and gray regions represent the surface layer ($l_1$) and the bulk material, respectively. During $t_1 \to t_2$, the laser polarization is aligned with the $y$ axis, yielding a polarization angle of $\theta_1=0^\circ$. $\beta$ is the angle between the material and laboratory coordinates. (c) and (d) Side views of the unperturbed $T_d-$WTe$_2$ in the $y'z'$ and $x'z'$ planes, respectively. (e) A representative model of the perturbed $T_d$-WTe$_2$ at $t_2=\Delta T$ s, whose susceptibility tensor is close to $e^{A_1(\theta_1+\beta)\Delta T}\cdot\chi^{(2)}_i$. (f) and (g) Side views of the perturbed material in the $y'z'$ and $x'z'$ planes, respectively, with small displacements ($\Delta y'$ and $\Delta x'$) of the surface layer.}
    \label{fig:dft}
\end{figure}

For future investigations, the existing evolution operator $e^{A(\theta +\beta)t}$ derived from the dynamic model (Eq.~\eqref{eAt}) can offers a bridge between characterization and first-principles calculations, thereby enabling identify the stimulus-induced intermediate states in materials. Specifically, characterization would establish which factor dominates the change (e.g., oxidation, defect formation, and phase transformation), corresponding structures can then be constructed for each candidate, and first-principles calculations could fix which of them is reasonable. To illustrate this, since characterization experiments are beyond the scope of this work, we assume that the phase transition is the main factor changing the material properties, as reported previously \cite{phase_transition1,phase_transition2,phase_transition3}. First, we can use density functional theory (DFT) to calculate $\chi^{(2)}$ for unperturbed $T_d-$WTe$_2$ (denoted as $\chi^{(2)}_i$), then substitute $\chi^{(2)}_i$ into the dynamic model and fit the test data to obtain the direction angle $\beta$ and dynamic parameters $a$ and $b$, and constructing the evolution operator $e^{A(\theta+\beta)t}$. To find the intermediate states, we take the first SHG test point (the red point in Fig.~\ref{fig:SHG}d) as an example, the process at this point is shown in Fig.~\ref{fig:dft}. As shown in Fig.~\ref{fig:dft}a, during the interval $t_1\to t_2$, $T_d-$WTe$_2$ is unperturbed at $t_1=0$ (Fig.~\ref{fig:dft}b and \ref{fig:dft}c), and its susceptibility tensor is $\chi^{(2)}_i$. Since the laser polarization is aligned with the $y$ axis (Fig.~\ref{fig:dft}b), we have $\theta_1=0^\circ$. Thus, the evolution operator $e^{A(\theta_1+\beta)t}$ transforms $\chi^{(2)}_i$ into $e^{A(\theta_1+\beta)\Delta T}\cdot\chi^{(2)}_i$ after $\Delta T$ seconds (i.e., at $t_2$, Fig.~\ref{fig:dft}a).   

Within the proposed framework, DFT calculations can provide a specific state and its susceptibility tensor close to $e^{A(\theta_1+\beta)\Delta T}\cdot\chi^{(2)}_i$ (Fig.~\ref{fig:dft}e-\ref{fig:dft}g). This state can be regarded as a suitable intermediate state. To investigate other intermediate states, the same workflow can be repeated by calculating $\prod_{i=1}^n e^{A_i(\theta_i+\beta)\Delta T} \cdot \chi^{(2)}_i$ and selecting a suitable structure with susceptibility tensor closely matches the calculated one.

\section{Conclusions and Discussion}
In this work, we investigate the dynamic properties of materials under various stimuli. We build dynamic models and verify the discrete model with an experiment. To study dynamic problems, we first establish a generalized ICO approach for constructing spatial operation matrices for material property coefficient matrices across diverse physical systems. By determining only the orders of the input and output vectors, the mathematical expressions of both the coefficient matrix and its corresponding operation matrix can be constructed. Based on this theoretical framework, we develop a Python program (the ``$\chi$ Program'') that enables researchers to conveniently handle high-order physical processes.

Under the ICO formalism, to create dynamic models, the stimulus ($\vec{E}$) applied to the $i$-th order properties of a material can be expressed as:
\begin{equation}
    O^{\otimes i}\cdot \vec{E}^{\otimes i}.
\end{equation}
Here, $O$ is a spatial operation matrix. When considering dynamic problems, it is necessary to calculate the derivative of the function with respect to $\alpha$ (a stimulus variable). The terms $O^{\otimes i}$ and $E^{\otimes i}$ can be easily differentiated as:
\begin{equation}
    \frac{d}{d\alpha}O^{\otimes i}=\sum_{k=1}^i O^{\otimes (k-1)}\otimes \frac{d O}{d\alpha} \otimes O^{\otimes (i-k)},
\end{equation}
where $O^{\otimes 0}=1$. In addition, for the continuous case (Section \ref{Continue model}), the solution of the dynamic equation is a Peano-Baker series, which is difficult to solve analytically. However, we can use the discrete model (Section \ref{Discrete Model}) to approximately solve this series by setting $\Delta t$ and $\theta_n-\theta_{n-1}$ to be sufficiently small.

Based on the ICO formalism and the dynamic model, we propose a new point-to-point approach for fitting SHG experimental data. By introducing $\frac{d \chi}{dt}$, the dynamic model captures the asymmetric character of the experimental data, whereas the conventional model is constrained to a symmetric curve. The fitting results show that our model performs better than the conventional one because it incorporates variables that characterize the behavior of the material during the experiment. Under practical conditions, additional factors (e.g., oxidation, defect generation, and phase transition) may influence the material. In future research, more characterization tests (e.g., scanning electron microscopy, transmission electron microscopy, and X-ray diffraction) should be performed to determine which factor predominantly affects the dynamic properties of materials. Accordingly, Eq.~\eqref{decay_chi} can be modified to account for these factors. Furthermore, after extracting the parameters of the material's dynamic properties, researchers can perform DFT calculations to investigate the mechanisms that affect material behavior.

Beyond bridging experiments and DFT calculations, the evolution operator $e^{A(\theta+\beta)t}$ of the dynamic model also assists theoretical research. By applying this operator to the susceptibility tensor, some elements lose the rigid relationships constrained by the pristine material symmetry, indicating that the external-field-induced symmetry breaking, which is currently a hot research topic.     

Therefore, this work provides an applicable solution for creating dynamic functions to investigate the interaction between fields and matter. We also propose a generalized ICO approach to simplify the calculation of spatial operations in various physical systems, a concise notation for understanding complex spatial operation processes, and a mathematical framework for solving dynamic problems in theoretical and experimental research.

\section*{Method}
\subsection*{Sample preparation}
The $T_d-$WTe$_2$ sheet were prepared by mechanical exfoliation and transferred onto a silicon substrate by a 2D material transfer platform.
\subsection*{SHG experiment}
SHG measurements were performed using a WITec Raman spectrometer. The excitation wavelength was 1064 nm, with a 600 g/mm grating (blaze wavelength: 500 nm). A Zeiss EC Epiplan-Neofluar 50$\times$/0.75
objective was employed, and a half-wave plate was placed between the filter and the objective lens. The polarization configuration was set to the parallel mode. The sample was mounted on a translation stage, and SHG intensities were recorded as a function of the half‑wave plate rotation angle in steps of $10^\circ$.

\section*{Data Availability}
The data in the Article are available from the corresponding author upon reasonable request.

\section*{Code Availability}
The ``$\chi$ Package'' is available through GitHub: \url{https://github.com/HongjinXiong/Chi_Package/tree/main}

\section*{Author Contributions}
T.M. proposed and supervised the project. H.X. conceived the project, performed the theoretical analysis, developed the code, conducted the data analysis, and drafted the manuscript. Z.X. carried out the SHG experiments. X.Y. supervised the theoretical aspects and contributed to manuscript revision. All the authors discussed the results and commented on the manuscript.

\section*{Competing Interests}
The authors declare no competing interests.

\section*{Acknowledgments}
T.M. acknowledges the funding support from the Innovation and Technology Support Programme of the Innovation and Technology Fund by the Innovation and Technology Commission (Grant No. ITS/150/23FP), the Research Grants Council of the Hong Kong Special Administrative Region, China (Project No. PolyU 15306824), and Start-up Fund for RAPs under the Strategic Hiring Scheme of The Hong Kong Polytechnic University (Grant No. P0042991).

Any opinions, findings, conclusions or recommendations expressed in this material/event (or by members of the project team) do not reflect the views of the Government of the Hong Kong Special Administrative Region, the Innovation and Technology Commission or the Panel of Assessors for the Innovation and Technology Support Programme of the Innovation and Technology Fund.

\bibliography{Main}

\clearpage

\appendix
\setcounter{equation}{0} 
\setcounter{figure}{0} 
\renewcommand{\theequation}{S.\arabic{equation}}
\renewcommand{\thefigure}{S\arabic{figure}}
\begin{center}
    {\Large Supplementary Materials of}
\end{center}
\vspace{0.2cm}
\begin{center}
{\Large\textbf{Dynamic Behaviors of Materials Under Angle-Resolved Experiments: A Matrix Framework for Nonlinear Physics}}
\end{center}

\section{Construct $O^{\otimes j}$}\label{construct}
We construct the $j$-th order spatial operation matrix $O^{\otimes j}$ for the three fundamental crystallographic spatial operations, including rotation, inversion, and mirror reflection. 

\subsubsection{Rotation}
For rotation,
\begin{align}
R_x(\phi_x) &= \left(\begin{matrix}
1 & 0&0 \\
0 & \cos(\phi_x)& -\sin (\phi_x)\\
0 & \sin(\phi_x)& \cos (\phi_x)
\end{matrix}\right), &
R_y(\phi_y) &= \left(\begin{matrix}
\cos(\phi_y) & 0&\sin(\phi_y) \\
0 & 1& 0\\
-\sin(\phi_y) & 0& \cos (\phi_y)
\end{matrix}\right), &
R_z(\phi_z) &= \left(\begin{matrix}
\cos(\phi_z) & -\sin(\phi_z)&0 \\
\sin(\phi_z)& \cos(\phi_z)& 0\\
0 & 0& 1
\end{matrix}\right),\label{RxRyRz}
\end{align}
are rotation matrices, and can rotate $\vec{J}$ and $\vec{E}$ around $x,y,z$ axes, respectively. The general rotation matrix is:
\begin{equation}
    R(\phi_x,~\phi_y,~\phi_z)=R_x(\phi_x)\cdot R_y(\phi_y)\cdot R_z(\phi_z),\label{R}
\end{equation}
where $\phi_x$, $\phi_y$ and $\phi_z$ represent the rotation angle around $x,~y,~z$ axes, respectively. And $R(\phi_x,~\phi_y,~\phi_z)$ can rotate a $3\times1$ vector to any direction. Follow the Eq.~(\ref{high_order_relation_operation_expanded}) and applying the rule $(A\cdot B)\otimes (A\cdot B)=(A\otimes A)\cdot (B\otimes B)$, we construct $j$-th order rotation matrices $R^{\otimes j}(\phi_x,~\phi_y,~\phi_z)$, $R_x^{\otimes j}(\phi_x)$, $R_y^{\otimes j}(\phi_y)$ and $R_z^{\otimes j}(\phi_z)$ for $\vec{J}^{\otimes j}$ and $\vec{E}^{\otimes j}$:
\begin{equation}
    R^{\otimes j}(\phi_x,~\phi_y,~\phi_z)=R_x^{\otimes j}(\phi_x)\cdot R_y^{\otimes j}(\phi_y)\cdot R_z^{\otimes j}(\phi_z).\label{H_R}
\end{equation}

The $R_i^{\otimes j}$ {\color{blue} ($i=x,~y,~z$)} has exactly the same excellent properties as the $R$:
\begin{gather}
       R_i^{\otimes j}(-\phi_i)=\big[R_i^{-1}(\phi_i)\big]^{\otimes j}=\big[R_i^{T}(\phi_i)\big]^{\otimes j};~~\big[R^{T}(\phi_x,~\phi_y,~\phi_z)\big]^{\otimes j}=(R^{-1})^{\otimes j};~~i=x,~y,~z.
\end{gather}

\subsubsection{Inversion}
Inversion is a central symmetry operation, in which the matrix $O=I$ enables $\vec{J}$ and $\vec{E}$ to change to $-\vec{J}$ and $-\vec{E}$, respectively. 
Thus, the inversion matrix: 
\begin{equation}
    I=-\mathbb{I}_{3\times3},\label{I} 
\end{equation}
is the $3\times 3$ negative identity matrix. According to {\color{blue}Eq.~(\ref{high_order_relation_operation_expanded})}, the inversion matrix for $\vec{J}^{\otimes j}$ and $\vec{E}^{\otimes j}$ is:
\begin{equation}
    I^{\otimes j}=(-\mathbb{I}_{3\times3})^{\otimes j}=(-1)^j \cdot \mathbb{I}_{3^j\times 3^j}\label{high_order_I}
\end{equation}

\subsubsection{Mirror reflection}
The basic mirror matrices are:
\begin{equation}
    M_{xy}=\left(\begin{array}{ccc}
        1 & 0 & 0 \\
        0 & 1 & 0 \\
        0 & 0 & -1 \\ 
    \end{array}\right);~
    M_{yz}=\left(\begin{array}{ccc}
        -1 & 0 & 0 \\
        0 & 1 & 0 \\
        0 & 0 & 1 \\ 
    \end{array}\right);~
    M_{zx}=\left(\begin{array}{ccc}
        1 & 0 & 0 \\
        0 & -1 & 0 \\
        0 & 0 & 1 \\ 
    \end{array}\right),
    \label{Mxyyzzx}
\end{equation}
with functions of mirroring $\vec{J}$ and $\vec{E}$ for $xy$, $yz$ and $xz$ planes, respectively. The connection between mirror matrices and inversion matrices (Eq.~(\ref{I})) is:
\begin{equation}
    I=M_{xy}\cdot M_{yz} \cdot M_{zx}.
\end{equation}

Identically, using the same method as in previous sections, the mirror matrices for $\vec{J}^{\otimes j}$ and $\vec{E}^{\otimes j}$ is {\color{blue} $M^{\otimes j}_i~(i=xy,~yz,~zx)$}, with properties:
\begin{gather}
    I^{\otimes j}=M_{xy}^{\otimes j}\cdot M_{yz}^{\otimes j} \cdot M_{zx}^{\otimes j};~~M_{i}^{\otimes j}=[M_{i}^{T}]^{\otimes j};~~i=xy,~yz,~zx.\label{H_M}
\end{gather}
Here, the $I^{\otimes j}$ is the inversion matrix (Eq.~(\ref{high_order_I})).

\section{Examine The Coefficient Matrix For Various Systems Through ICO Formalism }\label{Validation_sm}

To verify the ICO formalism developed in the previous section, we calculate the coefficient matrices for 2nd-order stimulated/1st-order response systems and  2nd-order stimulated/response systems, respectively. We demonstrate that these matrices recover the results in the literature. Additionally, we also use the 
``$\chi$ program" to resolve several conflicting results found in the literature.

\subsection{Second-order stimulate and first-order response system}\label{3m_symmetry}

In the following, we consider both a centrosymmetric crystal and a trigonal pyramid in the second-order stimulated and first-order response nonlinear optics.

For a centrosymmetric crystal, which is invariant under the inversion operator $I$, Eq.~\eqref{dede} holds with $m=1$, $n=2$, and $O=I$. Thus, we have:
\begin{equation}
    \chi^{(2)}=I\cdot\chi^{(2)}\cdot I^{\otimes 2}
    =-\chi^{(2)} ,
    \label{equation1}
\end{equation}
where we use the fact that $I$ is the negative of the identity matrix, as defined in Section~\ref{construct} of the SM. Therefore, $\chi^{(2)}=0$ is the solution of Eq.~\eqref{equation1}, which agrees with the result that centrosymmetric crystals have no second-order nonlinear optical response \cite{Nonlinear_optics,newnham,Physical_properties_of_crystals}.

\begin{figure}[h]
    \centering
    \includegraphics[width=0.6\linewidth]{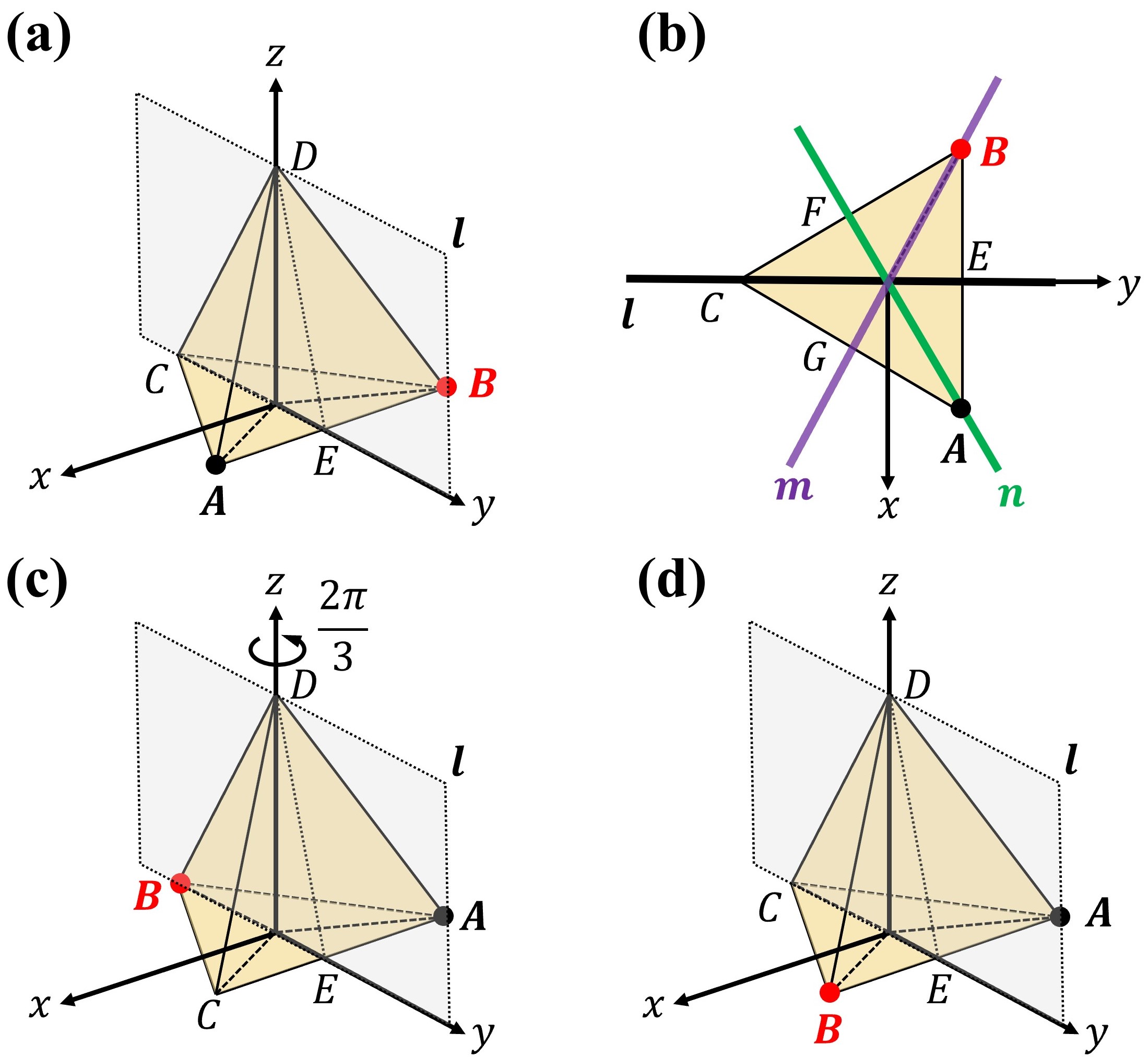}
    \caption{Trigonal pyramid. 
(a) A trigonal pyramid. (b) Top view of the trigonal pyramid. The three mirror planes $l$, $m$, and $n$ are indicated by the black, purple, and green lines, respectively. (c) The object is rotated counterclockwise by $\frac{2\pi}{3}$ about the $z$-axis. (d) The object is reflected about the $l$ plane.}
    \label{fig:trigonal_pyramid}
\end{figure}

For a trigonal pyramid with $3m$ symmetry, as shown in Fig.~\ref{fig:trigonal_pyramid}, the base is a regular triangle, namely plane $ABC$ in Fig.~\ref{fig:trigonal_pyramid}a. The pyramid has three mirror planes, denoted as the $n$, $l$, and $m$ planes in Fig.~\ref{fig:trigonal_pyramid}b. 
Without loss of generality, we choose the mirror plane $l$ to be aligned with the $yz$ plane, following the same choice as in Refs.~\cite{Physical_properties_of_crystals,About_the_calculation_of_the_second-order_susceptibility}. The coefficient matrix $\chi^{(1\times 2)}$ is a $3\times 9$ matrix, whose components are denoted by $\chi_{ijk}$, where $i,j,k$ take values from the set $(x,y,z)$. Because the trigonal pyramid has $C_3$ symmetry, it is invariant under a rotation of $\frac{2\pi}{3}$ about the $z$-axis, as shown in Fig.~\ref{fig:trigonal_pyramid}c. It is also invariant under the mirror operation with respect to the $l$-plane, as shown in Fig.~\ref{fig:trigonal_pyramid}d. Therefore, the coefficient matrix $\chi^{(1\times 2)}$ satisfies the following two equations
\begin{equation}
    \chi^{(2)}-R_z\left(\frac{2\pi}{3}\right)\cdot \chi^{(2)}\cdot \left[ R_z^{-1}\left(\frac{2\pi}{3}\right)\right]^{\otimes 2}=0,
    \label{rotation_result}
\end{equation}
and
\begin{equation}
    \chi^{(2)}-M_{yz}\cdot \chi^{(2)}\cdot \left[M_{yz}^{-1}\right]^{\otimes 2}=0.
    \label{myz}
\end{equation}
Solving Eqs.~\eqref{rotation_result} and \eqref{myz} by the ``$\chi$ program'' gives
\begin{gather}
    \chi^{(2)}=
    \begin{pmatrix}
    0 & \chi_{xxy} & \chi_{xxz} & \chi_{xyx} & 0 & 0 & \chi_{xzx} & 0 & 0 \\
    \chi_{yxx} & 0 & 0 & 0 & \chi_{yyy} & \chi_{yyz} & 0 & \chi_{yzy} & 0 \\
    \chi_{zxx} & 0 & 0 & 0 & \chi_{zyy} & 0 & 0 & 0 & \chi_{zzz}
    \end{pmatrix},
    \label{reduced_chi}
\end{gather}
where the nonzero components satisfy the following relations:
\begin{gather}
    \chi_{xxy} = -\chi_{yyy},~
    \chi_{xxz} = \chi_{yyz},~
    \chi_{xyx} = -\chi_{yyy}, \nonumber \\
    \chi_{xzx} = \chi_{yzy},~
    \chi_{yxx} = -\chi_{yyy},~
    \chi_{zxx} = \chi_{zyy}.
    \label{relation}
\end{gather}
This result agrees with the effective matrix written in Voigt notation \cite{Nonlinear_optics}. In this notation, the last two indices of $\chi_{ijk}$ are combined into one index $l$, which takes values from 1 to 6. The first index $i$ takes values 1, 2, and 3 instead of $x,y,z$. The correspondence is:
\begin{gather}
    i\rightarrow 1,2,3, \nonumber\\
    jk\rightarrow l:~xx=1,~yy=2,~zz=3,~yz,zy=4,~zx,xz=5,~xy,yx=6.
    \label{notation}
\end{gather}
For example, the relation $\chi_{xxy}=-\chi_{yyy}$ in Eq.~\eqref{relation} can be written as $d_{16}=-d_{22}$. Using the same correspondence, matrix~\eqref{reduced_chi} becomes
\begin{gather}
    \chi_{il}=
    \begin{pmatrix}
    0 & 0 & 0 & 0 & \chi_{24} & -\chi_{22} \\
    -\chi_{22} & \chi_{22} & 0 & \chi_{24} & 0 & 0 \\
    \chi_{31} & \chi_{31} & \chi_{33} & 0 & 0 & 0
    \end{pmatrix},
\end{gather}
which is the same as the form reported in previous works, such as Page 45 of Ref.~\cite{Nonlinear_optics} and Ref.~\cite{Theeffectsofphasematching}. By applying Kleinman's rule, the element $\chi_{24}$ is equal to $\chi_{31}$ in the effective tensor. This method allows the coefficient matrix to be written in a more compact form. However, the cost is that it loses the convenience of applying symmetry operations directly in the original tensor form.

To verify the correctness of the ICO formalism, we further derive the non-vanishing elements of the second-order susceptibility matrix for four representative crystal systems: triclinic, monoclinic, orthorhombic, and tetragonal crystals by using the ``$\chi$ program''(see Sec. \ref{sec:verification} of the Supplemental Material).

\subsection{Second-order stimulate and response system}\label{Other physical systems}
Our method can also be performed in other physical systems. We only need to analyze the order of input and output vector, then we can create the matrices to manipulate the coefficient matrix.

For example, in elasticity, the relation between stress ($\sigma$) and strain ($\epsilon$) is \cite{Physical_properties_of_crystals,newnham}:
\begin{equation}
    \sigma=C\cdot\epsilon\to \begin{pmatrix}
\sigma_1 \\
\sigma_2 \\
\sigma_3 \\
\sigma_4 \\
\sigma_5 \\
\sigma_6
\end{pmatrix}
=
\begin{pmatrix}
C_{11} & C_{12} & C_{13} & C_{14} & C_{15} & C_{16} \\
C_{21} & C_{22} & C_{23} & C_{24} & C_{25} & C_{26} \\
C_{31} & C_{32} & C_{33} & C_{34} & C_{35} & C_{36} \\
C_{41} & C_{42} & C_{43} & C_{44} & C_{45} & C_{46} \\
C_{51} & C_{52} & C_{53} & C_{54} & C_{55} & C_{56} \\
C_{61} & C_{62} & C_{63} & C_{64} & C_{65} & C_{66}
\end{pmatrix}
\begin{pmatrix}
\epsilon_1 \\
\epsilon_2 \\
\epsilon_3 \\
\epsilon_4 \\
\epsilon_5 \\
\epsilon_6
\end{pmatrix} \label{stress-strain}.
\end{equation}
The $C$ in Eq.~(\ref{stress-strain}) denotes the elasticity matrix. The components of $\sigma$, $C$ and $\epsilon$ are expressed in Voigt notation (Eq.~(\ref{notation})). However, this notation complicates matrix manipulation. Therefore, to demonstrate the advantage of ICO formalism, we first rewrite $\sigma$ and $\epsilon$ as follows:
\begin{align}
 \sigma&\equiv\begin{pmatrix}
\sigma_{xx} ,\sigma_{xy}, \sigma_{xz} ,\sigma_{yx},\sigma_{yy}, \sigma_{yz},\sigma_{zx},\sigma_{zy},\sigma_{zz}
\end{pmatrix}^T,\nonumber\\
\epsilon&\equiv\begin{pmatrix}
\epsilon_{xx},\epsilon_{xy},\epsilon_{xz},\epsilon_{yx},\epsilon_{yy},\epsilon_{yz},\epsilon_{zx},\epsilon_{zy},\epsilon_{zz}
\end{pmatrix}^T. \label{unreduced-stress-strain}
\end{align}

The components $\sigma_{ab},~(a,b=x,y,z)$ in $\sigma$ can be expressed as \cite{stress_strain}:
\begin{equation}
    \sigma_{ab} =\sigma_{ba}=  \frac{ F_a}{ S_b},\label{physical_meaning_of_sigma}
\end{equation}
where $F_a$ is the component of force vector $\vec{F}=(F_x,F_y,F_z)$, and $S_b$ is the component of area vector $\vec{S}=(S_x,S_y,S_z)$. Thus, to apply ICO formalism, we decompose the $\sigma$ in Eq.~(\ref{unreduced-stress-strain}) as:
\begin{equation}
    \sigma
\equiv\begin{pmatrix}
F_x \\
F_y\\
F_z
\end{pmatrix}
\otimes
\begin{pmatrix}
\frac{1}{ S_x} \\[6pt]
\frac{1}{ S_y}\\[6pt]
\frac{1}{ S_z}
\end{pmatrix}=\vec{F}\otimes \vec{S}\label{sigma_expression}.
\end{equation}
Identically, for the components $\epsilon_{ab},~(a,b=x,y,z)$ in $\epsilon$ (Eq.~(\ref{unreduced-stress-strain})) can be express as \cite{Physical_properties_of_crystals}:
\begin{equation}
    \epsilon_{ab} = \frac{1}{2}\lim_{ \Delta \to 0}\bigg( \frac{\Delta u_a}{\Delta b}+\frac{\Delta u_b}{\Delta a}\bigg),\label{physical_meaning_of_epsilon}
\end{equation}
where $u$ represents the elongation of materials. To apply ICO formalism, we express $\epsilon$ in Eq.~(\ref{unreduced-stress-strain}) as:
\begin{equation}
    \epsilon
\equiv\frac{1}{2}\lim_{\Delta\to0}\left[
\begin{pmatrix}
\Delta u_x \\
\Delta u_y\\
\Delta u_z
\end{pmatrix}
\otimes
\begin{pmatrix}
\frac{1}{\Delta x} \\[6pt]
\frac{1}{\Delta y}\\[6pt]
\frac{1}{\Delta z}
\end{pmatrix}+
\begin{pmatrix}
\frac{1}{\Delta x} \\[6pt]
\frac{1}{\Delta y}\\[6pt]
\frac{1}{\Delta z}
\end{pmatrix}\otimes \begin{pmatrix}
\Delta u_x \\
\Delta u_y\\
\Delta u_z
\end{pmatrix}\right]
=\frac{1}{2}\big(\vec{u}\otimes \vec{X}+\vec{X}\otimes \vec{u}\big)\label{epsilon_expression}.
\end{equation}

Thus, both $\sigma$ and $\epsilon$ are second-order variables, and the corresponding expression to operate them is $O^{\otimes 2}$ (where $O$ the spatial operation matrix in three dimensional space) because (refer to Eq.\eqref{sigma_expression} and Eq.\eqref{epsilon_expression}):
\begin{align}
    O^{\otimes2}\cdot \sigma&
=(O\cdot\vec{F})\otimes (O\cdot\vec{S}),\nonumber\\
O^{\otimes2}\cdot \epsilon&=\frac{1}{2}\big[(O\cdot\vec{u})\otimes (O\cdot\vec{X})+(O\cdot\vec{X})\otimes (O\cdot\vec{u})\big].
\end{align}
Therefore, according to Eq.\eqref{O_chi}, the mathematical expression to transform the elasticity matrix $C$ is:
\begin{equation}
    O^{\otimes 2}\cdot C\cdot (O^{-1})^{\otimes 2}\label{C_control}.
\end{equation}

The $O^{\otimes 2}$ matrix can be applied not only in elasticity but also to other systems, such as electrostriction and magnetostriction, since such systems share the same mathematical formulation as elasticity. According to Ref. \cite{newnham}, the relation between strain $\epsilon$ and electric field $\vec{E}$ (electrostriction) is $\epsilon=M\cdot \vec{E}\vec{E}$, and the relation between and strain ($\epsilon$) and magnetization $\vec{I}$ (magnetostriction) is $\epsilon=N\cdot \vec{I}\vec{I}$. The terms $\epsilon$, $\vec{E}\vec{E}$, $\vec{I}\vec{I}$ are second-order variables and can be processed by $O^{\otimes 2}$ for spatial operations. Therefore, the mathematical expressions of operating on the coefficient matrices $M$ and $N$ are the same as that in Eq.\eqref{C_control}.

To verify the correctness of Eq.~(\ref{C_control}), we use it to derive the relation between the elements of coefficient matrices of elasticity, electrostriction and magnetostriction under the material with $C_3$ symmetry, and the results align with
previous works (see section \ref{other physical systems_SM}, SM). 

Therefore, our formalism enables the construction of operation matrices that act on higher-order vectors and higher-dimensional coefficient matrices in various physical systems. Furthermore, using the ``$\chi$ program'', we find that for $C_3$ crystals the magnetostriction matrix element satisfies $N_{45} = -N_{54}$, in agreement with Eq. (92) of Ref. \cite{magnetostriction}. This contradicts the ``$N_{45} = N_{54}$'' reported in Table 15.5 of Ref. \cite{newnham}. For both $C_3$ and $3m$ crystals, the piezoelectric coefficients are found to satisfy $d_{16} = -d_{22}$, which is consistent with Eq. (A5) of Ref. \cite{C3_piezoelectric_tensor_violate} ($C_3$) and Fig. 2 of Ref. \cite{3m_piezoelectric_tensor_violate} ($3m$). This indicates that the ``$d_{16} = -2d_{22}$'' reported in Table 8 of Ref. \cite{Physical_properties_of_crystals} and Table 12.2 of Ref. \cite{newnham} is incorrect.

\section{Verification}\label{sec:verification}
We using our method and the ``$\chi$ program'' (Section \ref{ICO_chi}) to obtain nonvanishing elements of susceptibility tensors for $4$ crystal systems, which corresponds to previous results (Nonlinear optics pages $46$, Table 1.5.2 \cite{Nonlinear_optics}). In this section, we don't consider the Kleinman's rule, the $\chi^{(2)}$ is:
\begin{gather}
    \chi^{(2)}=    \begin{pmatrix}
\chi_{xxx} & \chi_{xxy} & \chi_{xxz} & \chi_{xyx} & \chi_{xyy} & \chi_{xyz} &\chi_{xzx} &\chi_{xzy} &\chi_{xzz} \\
\chi_{yxx} & \chi_{yxy} & \chi_{yxz} & \chi_{yyx} & \chi_{yyy} & \chi_{yyz} &\chi_{yzx} &\chi_{yzy} &\chi_{yzz} \\
\chi_{zxx} & \chi_{zxy} & \chi_{zxz} & \chi_{zyx} & \chi_{zyy} & \chi_{zyz} &\chi_{zzx} &\chi_{zzy} &\chi_{zzz}
\end{pmatrix}.
\end{gather}

\subsubsection{Triclinic}
\begin{center}
    1. Class $1=C_1$
\end{center}

For triclinic crystal belonging to class $1=C_1$, refer to Eq.~(\ref{RxRyRz}), Eq.~(\ref{H_R}) and Eq.~(\ref{dede}), we have:
\begin{equation}
    \chi^{(2)}=R(2\pi,2\pi,2\pi)\cdot\chi^{(2)}\cdot [R^{T}(2\pi,2\pi,2\pi)]^{\otimes 2}=\chi^{(2)}\label{triclinic_1=C1},
\end{equation}
which means all elements in $\chi^{(2)}$ are independent and nonzero.
\\\\
\begin{center}
    2. Class $\overline{1}=S_2$
\end{center}

For triclinic crystal belonging to class $\overline{1}=S_2$, refer to Eq.~(\ref{I}), Eq.~(\ref{high_order_I}) and Eq.~(\ref{dede}), we have:
\begin{equation}
    \chi^{(2)}=I\cdot\chi^{(2)}\cdot [I^{-1}]^{\otimes 2}\to\chi^{(2)}=0\label{triclinic_1-=S2},
\end{equation}
which means all elements in $\chi^{(2)}$ vanish.

\subsubsection{Monoclinic}
\begin{center}
    1. Class $2=C_2$
\end{center}

For monoclinic crystal belonging to class $2=C_2$, and the $2$-fold rotation axis parallel to $y$ axis, refer to Eq.~(\ref{RxRyRz}), Eq.~(\ref{H_R}) and Eq.~(\ref{dede}), we have:
\begin{equation}
    \chi^{(2)}=R_y(\pi)\cdot\chi^{(2)}\cdot [{R_y^{T}}(\pi)]^{\otimes 2},\label{monoclinic_2y}
\end{equation}
which gives:
\begin{gather}
    \chi_{xxx} = 0,~
\chi_{xxz} = 0,~
\chi_{xyy} = 0,~
\chi_{xzx} = 0,~
\chi_{xzz} = 0,~
\chi_{yxy} = 0,~
\chi_{yyx} = 0,~\nonumber\\
\chi_{yyz} = 0,~
\chi_{yzy} = 0,~
\chi_{zxx} = 0,~
\chi_{zxz} = 0,~
\chi_{zyy} = 0,~
\chi_{zzx} = 0,~
\chi_{zzz} = 0.
\end{gather}
\\\\

\begin{center}
    2. Class $m=C_{1h}$
\end{center}

For monoclinic crystal belonging to class $m=C_{1h}$, and the mirror plane perpendicular to $y$ axis ($zx$ plane), refer to Eq.~(\ref{Mxyyzzx}), Eq.~(\ref{H_M}) and Eq.~(\ref{dede}), we have:
\begin{equation}
    \chi^{(2)}=M_{zx}\cdot \chi^{(2)} \cdot [{M_{zx}^{-1}}]^{\otimes 2},
\end{equation}
which gives:
\begin{gather}
    \chi_{xxy} = 0,~
\chi_{xyx} = 0,~
\chi_{xyz} = 0,~
\chi_{xzy} = 0,~
\chi_{yxx} = 0,~
\chi_{yxz} = 0,~
\chi_{yyy} = 0,~\nonumber\\
\chi_{yzx} = 0,~
\chi_{yzz} = 0,~
\chi_{zxy} = 0,~
\chi_{zyx} = 0,~
\chi_{zyz} = 0,~
\chi_{zzy} = 0.
\end{gather}
\\\\
\begin{center}
    3. Class $2/m=C_{2h}$
\end{center}

For monoclinic crystal belonging to class $2/m=C_{2h}$, there is a $2$-fold axis in any direction and a mirror plane perpendicular to this axis. By rotating the object, the $2$-fold axis can be $z$ axis and the mirror plane becomes $xy$ plane. Refer to Eq.~(\ref{RxRyRz}), Eq.~(\ref{H_R}), Eq.~(\ref{Mxyyzzx}), Eq.~(\ref{H_M}) and Eq.~(\ref{dede}), we have:
\begin{gather}
    \chi^{(2)}=R_z(\pi)\cdot\chi^{(2)}\cdot [{R_z^{T}}(\pi)]^{\otimes 2},~~\chi^{(2)}=M_{xy}\cdot \chi^{(2)}\cdot [{M_{xy}^{-1}}]^{\otimes 2}.
\end{gather}
Solving the equations, we have:
\begin{gather}
    \chi_{xxx} = 0,~
\chi_{xxy} = 0,~
\chi_{xxz} = 0,~
\chi_{xyx} = 0,~
\chi_{xyy} = 0,~
\chi_{xyz} = 0,~
\chi_{xzx} = 0,~
\chi_{xzy} = 0,~
\chi_{xzz} = 0,~
\chi_{yxx} = 0,~\nonumber\\
\chi_{yxy} = 0,~
\chi_{yxz} = 0,~
\chi_{yyx} = 0,~
\chi_{yyy} = 0,~
\chi_{yyz} = 0,~
\chi_{yzx} = 0,~
\chi_{yzy} = 0,~
\chi_{yzz} = 0,~
\chi_{zxx} = 0,~
\chi_{zxy} = 0,~\nonumber\\
\chi_{zxz} = 0,~
\chi_{zyx} = 0,~
\chi_{zyy} = 0,~
\chi_{zyz} = 0,~
\chi_{zzx} = 0,~
\chi_{zzy} = 0,~
\chi_{zzz} = 0.
\end{gather}
Thus, all elements in $\chi^{(2)}$ vanishing.

\subsubsection{Orthorhombic}
\begin{center}
    1. Class $222=D_2$
\end{center}

 For Orthorhombic crystal belonging to class $222=D_2$, refer to Eq.~(\ref{R}), Eq.~(\ref{H_R}) and Eq.~(\ref{dede}), we have:
\begin{gather}
    \chi^{(2)}=R_x(\pi)\cdot\chi^{(2)}\cdot [{R_x}^{T}(\pi)]^{\otimes 2},~~
    \chi^{(2)}=R_y(\pi)\cdot\chi^{(2)}\cdot [{R_y}^{T}(\pi)]^{\otimes 2},~~
    \chi^{(2)}=R_z(\pi)\cdot\chi^{(2)}\cdot [{R_z}^{T}(\pi)]^{\otimes 2},
\end{gather}
which gives:
\begin{gather}
\chi_{xxx} = 0,~
\chi_{xxy} = 0,~
\chi_{xxz} = 0,~
\chi_{xyx} = 0,~
\chi_{xyy} = 0,~
\chi_{xzx} = 0,~
\chi_{xzz} = 0,~
\chi_{yxx} = 0,~
\chi_{yxy} = 0,~
\chi_{yyx} = 0,~\nonumber\\
\chi_{yyy} = 0,~
\chi_{yyz} = 0,~
\chi_{yzy} = 0,~
\chi_{yzz} = 0,~
\chi_{zxx} = 0,~
\chi_{zxz} = 0,~
\chi_{zyy} = 0,~
\chi_{zyz} = 0,~
\chi_{zzx} = 0,~
\chi_{zzy} = 0,~\nonumber\\
\chi_{zzz} = 0.
\end{gather}
\\\\
\begin{center}
    2. Class $mm2=C_{2v}$
\end{center}

For Orthorhombic crystal belonging to class $mm2=C_{2v}$, refer to Eq.~(\ref{RxRyRz}), Eq.~(\ref{H_R}), Eq.~(\ref{Mxyyzzx}), Eq.~(\ref{H_M}) and Eq.~(\ref{dede}), we have:
\begin{gather}
    \chi^{(2)}=M_{yz}\cdot\chi^{(2)}\cdot [{M_{yz}^{-1}}]^{\otimes 2},~~\chi^{(2)}=M_{zx}\cdot\chi^{(2)}\cdot [{M_{zx}^{-1}}]^{\otimes 2},~~
     \chi^{(2)}=R_z(\pi)\cdot\chi^{(2)}\cdot [{R_z^{T}}(\pi)]^{\otimes 2},
\end{gather}
which gives:
\begin{gather}
    \chi_{xxx} = 0,~
\chi_{xxy} = 0,~
\chi_{xyx} = 0,~
\chi_{xyy} = 0,~
\chi_{xyz} = 0,~
\chi_{xzy} = 0,~
\chi_{xzz} = 0,~
\chi_{yxx} = 0,~
\chi_{yxy} = 0,~
\chi_{yxz} = 0,~\nonumber\\
\chi_{yyx} = 0,~
\chi_{yyy} = 0,~
\chi_{yzx} = 0,~
\chi_{yzz} = 0,~
\chi_{zxy} = 0,~
\chi_{zxz} = 0,~
\chi_{zyx} = 0,~
\chi_{zyz} = 0,~
\chi_{zzx} = 0,~
\chi_{zzy} = 0.
\end{gather}
\\\\
\begin{center}
    3. Class $mmm=D_{2h}$
\end{center}

For Orthorhombic crystal belonging to class $mmm=D_{2h}$, refer to Eq.~(\ref{Mxyyzzx}), Eq.~(\ref{H_M}) and Eq.~(\ref{dede}), we have:
\begin{align}
    \chi^{(2)}=M_{yz}\cdot\chi^{(2)}\cdot [{M_{yz}^{-1}}]^{\otimes 2},~~
    \chi^{(2)}=M_{zx}\cdot\chi^{(2)}\cdot [{M_{zx}^{-1}}]^{\otimes 2},~~
     \chi^{(2)}=M_{xy}\cdot\chi^{(2)}\cdot [{M_{xy}^{-1}}]^{\otimes 2},
\end{align}
which means:
\begin{equation}
    \chi^{(2)}=0.
\end{equation}
All elements of $\chi^{(2)}$ vanishing.

\subsubsection{Tetragonal}
\begin{center}
    1. Class $4=C_4$
\end{center}

For tetragonal crystal belonging to class $4=C_4$, the $4$-fold rotation axis is $z$ axis, refer to Eq.~(\ref{RxRyRz}), Eq.~(\ref{H_R}) and Eq.~(\ref{dede}), we have:
\begin{align}
    \chi^{(2)}&=R_z(\frac{\pi}{2})\cdot\chi^{(2)}\cdot [{R_z}^{T}(\frac{\pi}{2})]^{\otimes 2},
\end{align}
which gives:
\begin{gather}
    \chi_{xxx} = 0,~\chi_{xxy} = 0,~\chi_{xxz} = \chi_{yyz},~\chi_{xyx} = 0,~\chi_{xyy} = 0,~\chi_{xyz} = -\chi_{yxz},~\chi_{xzx} = \chi_{yzy},~\chi_{xzy} = -\chi_{yzx},~\nonumber\\
    \chi_{xzz} = 0,~
    \chi_{yxx} = 0,~
    \chi_{yxy} = 0,~
    \chi_{yyx} = 0,~ 
    \chi_{yyy} = 0,~
\chi_{yzz} = 0,~
\chi_{zxx} = \chi_{zyy},~
\chi_{zxy} = -\chi_{zyx},~
\chi_{zxz} = 0,\nonumber\\
\chi_{zyz} = 0,~
\chi_{zzx} = 0,~
\chi_{zzy} = 0.
\end{gather}
\\\\
\begin{center}
    2. Class $\overline{4}=S_4$
\end{center}

For tetragonal crystal belonging to class $\overline{4}=S_4$, the $4$-fold rotation axis is $z$ axis, refer to Eq.~(\ref{RxRyRz}), Eq.~(\ref{H_R}) and Eq.~(\ref{dede}), we have:
\begin{align}
    \chi^{(2)}&=R_z(\frac{\pi}{2})\cdot\chi^{(2)}\cdot [{R_z}^{T}(\frac{\pi}{2})]^{\otimes 2},
\end{align}
which gives:
\begin{gather}
    \chi_{xxx} = 0,~
\chi_{xxy} = 0,~
\chi_{xxz} = -\chi_{yyz},~
\chi_{xyx} = 0,~
\chi_{xyy} = 0,~
\chi_{xyz} = \chi_{yxz},~
\chi_{xzx} = -\chi_{yzy},~
\chi_{xzy} = \chi_{yzx},~
\chi_{xzz} = 0,~\nonumber\\
\chi_{yxx} = 0,~
\chi_{yxy} = 0,~
\chi_{yyx} = 0,~
\chi_{yyy} = 0,~
\chi_{yzz} = 0,~
\chi_{zxx} = -\chi_{zyy},~
\chi_{zxy} = \chi_{zyx},~
\chi_{zxz} = 0,~
\chi_{zyz} = 0,~
\chi_{zzx} = 0,~\nonumber\\
\chi_{zzy} = 0,~
\chi_{zzz} = 0.
\end{gather}
\\\\
\begin{center}
    3. Class $422=D_4$
\end{center}

For tetragonal crystal belonging to class $422=D_4$, the $4$-fold rotation axis is $z$ axis, and the $2$-fold rotation axes are $x$ and $y$ axes, respectively. Refer to Eq.~(\ref{RxRyRz}), Eq.~(\ref{H_R}) and Eq.~(\ref{dede}), we have:
\begin{align}
    \chi^{(2)}=R_x(\pi)\cdot\chi^{(2)}\cdot [{R_x}^{T}(\pi)]^{\otimes 2},~~
    \chi^{(2)}=R_y(\pi)\cdot\chi^{(2)}\cdot [{R_y}^{T}(\pi)]^{\otimes 2},~~
    \chi^{(2)}=R_z(\frac{\pi}{2})\cdot\chi^{(2)}\cdot [{R_z}^{T}(\frac{\pi}{2})]^{\otimes 2},
\end{align}
which gives:
\begin{gather}
   \chi_{xxx} = 0,~
\chi_{xxy} = 0,~
\chi_{xxz} = 0,~
\chi_{xyx} = 0,~
\chi_{xyy} = 0,~
\chi_{xyz} = -\chi_{yxz},~
\chi_{xzx} = 0,~
\chi_{xzy} = -\chi_{yzx},~\nonumber\\
\chi_{xzz} = 0,~
\chi_{yxx} = 0,~
\chi_{yxy} = 0,~
\chi_{yyx} = 0,~
\chi_{yyy} = 0,~
\chi_{yyz} = 0,~
\chi_{yzy} = 0,~
\chi_{yzz} = 0,~
\chi_{zxx} = 0,~\nonumber\\
\chi_{zxy} = -\chi_{zyx},~
\chi_{zxz} = 0,~
\chi_{zyy} = 0,~
\chi_{zyz} = 0,~
\chi_{zzx} = 0,~
\chi_{zzy} = 0,~
\chi_{zzz} = 0.
\end{gather}
\\
\begin{center}
    4. Class $4mm=C_{4v}$
\end{center}

For tetragonal crystal belonging to class $4mm=C_{4v}$, the $4$-fold rotation axis is $z$ axis, and mirror planes are $yz$ and $zx$ planes, respectively. Refer to Eq.~(\ref{RxRyRz}), Eq.~(\ref{H_R}), Eq.~(\ref{Mxyyzzx}), Eq.~(\ref{H_M}) and Eq.~(\ref{dede}), we have:
\begin{align}
    \chi^{(2)}=M_{yz}\cdot\chi^{(2)}\cdot [{M_{yz}^{-1}}]^{\otimes 2},~~
    \chi^{(2)}=M_{zx}\cdot\chi^{(2)}\cdot [{M_{zx}^{-1}}]^{\otimes 2},~~
    \chi^{(2)}=R_z(\frac{\pi}{2})\cdot\chi^{(2)}\cdot [{R_z}^{T}(\frac{\pi}{2})]^{\otimes 2},
\end{align}
which gives:
\begin{gather}
 \chi_{xxx} = 0,~
\chi_{xxy} = 0,~
\chi_{xxz} = \chi_{yyz},~
\chi_{xyx} = 0,~
\chi_{xyy} = 0,~
\chi_{xyz} = 0,~
\chi_{xzx} = \chi_{yzy},~
\chi_{xzy} = 0,~\nonumber\\
\chi_{xzz} = 0,~
\chi_{yxx} = 0,~
\chi_{yxy} = 0,~
\chi_{yxz} = 0,~
\chi_{yyx} = 0,~
\chi_{yyy} = 0,~
\chi_{yzx} = 0,~
\chi_{yzz} = 0,~\nonumber\\
\chi_{zxx} = \chi_{zyy},~
\chi_{zxy} = 0,~
\chi_{zxz} = 0,~
\chi_{zyx} = 0,~
\chi_{zyz} = 0,~
\chi_{zzx} = 0,~
\chi_{zzy} = 0.
\end{gather}
\\\\
\begin{center}
    5. Class $\overline{4}2m=D_{2d}$
\end{center}

For tetragonal crystal belonging to class $\overline{4}2m=D_{2d}$, the $4$-fold rotation axis is $z$ axis, the $2$-fold rotation axes are $x$ and $y$ axes, and mirror planes are $x=y$ and $x=-y$ planes. To calculate the mirror operation, we first rotate the object $\frac{\pi}{4}$ about $z$ axis, the two mirror planes align with $yz$ and $xz$ planes, respectively. Then, after mirror, by rotating the object $-\frac{\pi}{4}$ about $z$ axis, we finish the whole mirror operation. Refer to Eq.~(\ref{RxRyRz}), Eq.~(\ref{H_R}), Eq.~(\ref{Mxyyzzx}), Eq.~(\ref{H_M}) and Eq.~(\ref{dede}), we have:
\begin{align}
    \chi^{(2)}&=R_x(\pi)\cdot\chi^{(2)}\cdot [{R_x}^{T}(\pi)]^{\otimes 2}\nonumber\\
    \chi^{(2)}&=R_y(\pi)\cdot\chi^{(2)}\cdot [{R_y}^{T}(\pi)]^{\otimes 2}\nonumber\\
    \chi^{(2)}&=I\cdot R_z(\frac{\pi}{2})\cdot\chi^{(2)}\cdot [{R_z}^{T}(\frac{\pi}{2})]^{\otimes 2}\cdot [I^{\otimes 2}]^{-1}\nonumber\\
    \chi^{(2)}&=R_z(\frac{\pi}{4})^T\cdot M_{yz}\cdot R_z(\frac{\pi}{4})\cdot\chi^{(2)}\cdot [{R_z}^{T}(\frac{\pi}{4})]^{\otimes 2}\cdot [{M_{yz}^{-1}}]^{\otimes 2}\cdot [{R_z}^{\otimes 2}(\frac{\pi}{4})]\nonumber\\
    \chi^{(2)}&=R_z(\frac{\pi}{4})^T\cdot M_{zx}\cdot R_z(\frac{\pi}{4})\cdot\chi^{(2)}\cdot [{R_z}^{T}(\frac{\pi}{4})]^{\otimes 2}\cdot [{M_{zx}^{-1}}]^{\otimes 2}\cdot [{R_z}^{\otimes 2}(\frac{\pi}{4})],
\end{align}
which gives:
\begin{gather}
\chi_{xxx} = 0,~
\chi_{xxy} = 0,~
\chi_{xxz} = 0,~
\chi_{xyx} = 0,~
\chi_{xyy} = 0,~
\chi_{xyz} = \chi_{yxz},~
\chi_{xzx} = 0,~
\chi_{xzy} = \chi_{yzx},~\nonumber\\
\chi_{xzz} = 0,~
\chi_{yxx} = 0,~
\chi_{yxy} = 0,~
\chi_{yyx} = 0,~
\chi_{yyy} = 0,~
\chi_{yyz} = 0,~
\chi_{yzy} = 0,~
\chi_{yzz} = 0,~
\chi_{zxx} = 0,~\nonumber\\
\chi_{zxy} = \chi_{zyx},~
\chi_{zxz} = 0,~
\chi_{zyy} = 0,~
\chi_{zyz} = 0,~
\chi_{zzx} = 0,~
\chi_{zzy} = 0,~
\chi_{zzz} = 0.
\end{gather}
\\\\
\begin{center}
    6. Class $4/m=C_{4h}$ 
\end{center}

For tetragonal crystal belonging to class $4/m=C_{4h}$, the $4$-fold rotation axis is $z$ axis, and the mirror plane is $xy$ plane. Refer to Eq.~(\ref{RxRyRz}), Eq.~(\ref{H_R}), Eq.~(\ref{Mxyyzzx}), Eq.~(\ref{H_M}) and Eq.~(\ref{dede}), we have:
\begin{align}
    \chi^{(2)}=R_z(\frac{\pi}{2})\cdot\chi^{(2)}\cdot [{R_z}^{T}(\frac{\pi}{2})]^{\otimes 2},~~
    \chi^{(2)}=M_{xy}\cdot\chi^{(2)}\cdot [{M_{xy}^{-1}}]^{\otimes 2},\label{4/m=C_{4h}}
\end{align}
which gives:
\begin{gather}
\chi_{xxy} = 0,~
\chi_{xxz} = 0,~
\chi_{xyx} = 0,~
\chi_{xyy} = 0,~
\chi_{xyz} = 0,~
\chi_{xzx} = 0,~
\chi_{xzy} = 0,~
\chi_{xzz} = 0,~
\chi_{yxx} = 0,~
\chi_{yxy} = 0,~\nonumber\\
\chi_{yxz} = 0,~
\chi_{yyx} = 0,~
\chi_{yyy} = 0,~
\chi_{yyz} = 0,~
\chi_{yzx} = 0,~
\chi_{yzy} = 0,~
\chi_{yzz} = 0,~
\chi_{zxx} = 0,~
\chi_{zxy} = 0,~
\chi_{zxz} = 0,~\nonumber\\
\chi_{zyx} = 0,~
\chi_{zyy} = 0,~
\chi_{zyz} = 0,~
\chi_{zzx} = 0,~
\chi_{zzy} = 0,~
\chi_{zzz} = 0.
\end{gather}
\\
\begin{center}
    7. Class $4/mmm=D_{4h}$
\end{center}

For tetragonal crystal belonging to class $4/mmm=D_{4h}$, the $4$-fold rotation axis is $z$ axis, and the mirror planes are $xy$, $yz$ and $zx$ planes. Due to the derivation of No. 6 section (class $4/m=C_{4h}$), the condition ``$4$-fold rotation axis is $z$ axis, and the mirror planes are $xy$" can make the $\chi^{(2)}=0$. Thus, all elements in $\chi^{(2)}$ vanishing.

\section{Other Physical Systems}\label{other physical systems_SM}

As discussed in section \ref{Other physical systems}, in the field of elasticity, electrostriction and magnetostriction, since the order of input stimuli and out put signal are $2$, the coefficient matrices (refer as $N$) of these systems can be operated by:
\begin{gather}
    O^{\otimes 2}\cdot N\cdot [{O^{-1}}]^{\otimes 2}.
\end{gather}

To prove it, taking $C_3$ crystal as example. By rotating $\frac{2\pi}{3}$ around $z$ axis, the crystal should coincide with itself, thus refer to Eq.~(\ref{RxRyRz}), Eq.~(\ref{H_R}), Eq.~(\ref{dede}):
\begin{equation}
    N-\bigg[R_z(\frac{2\pi}{3})\bigg]^{\otimes 2}\cdot N \cdot \bigg[R_z^T(\frac{2\pi}{3})\bigg]^{\otimes 2}=0
\end{equation}
Solving this equation, we have:
\begin{gather}
N_{xxxx} = N_{yyyy},~
N_{xxxy} = -N_{yyyx},~
N_{xxxz} = -N_{yyxz},~
N_{xxyx} = -N_{yyxy},~
N_{xxyy} = N_{yyxx},~
N_{xxyz} = -N_{yyyz},~\nonumber\\
N_{xxzx} = -N_{yyzx},~
N_{xxzy} = -N_{yyzy},~
N_{xxzz} = N_{yyzz},~
N_{xyxx} = -N_{yxyy},~
N_{xyxy} = N_{yxyx},~
N_{xyxz} = -N_{yyyz},~\nonumber\\
N_{xyyx} = -N_{yxyx} - N_{yyxx} + N_{yyyy},~
N_{xyyy} = -N_{yxyy} - N_{yyxy} - N_{yyyx},~
N_{xyyz} = N_{yyxz},~
N_{xyzx} = -N_{yyzy},~\nonumber\\
N_{xyzy} = N_{yyzx},~
N_{xyzz} = -N_{yxzz},~
N_{xzxx} = -N_{yzyx},~
N_{xzxy} = -N_{yzyy},~
N_{xzxz} = N_{yzyz},~
N_{xzyx} = -N_{yzyy},~\nonumber\\
N_{xzyy} = N_{yzyx},~
N_{xzyz} = -N_{yzxz},~
N_{xzzx} = N_{yzzy},~
N_{xzzy} = -N_{yzzx},~
N_{xzzz} = 0,~\nonumber\\
N_{yxxx} = N_{yxyy} + N_{yyxy} + N_{yyyx},~
N_{yxxy} = -N_{yxyx} - N_{yyxx} + N_{yyyy},~
N_{yxxz} = -N_{yyyz},~
N_{yxyz} = N_{yyxz},~\nonumber\\
N_{yxzx} = -N_{yyzy},~
N_{yxzy} = N_{yyzx},~
N_{yzxx} = -N_{yzyy},~
N_{yzxy} = N_{yzyx},~
N_{yzzz} = 0,~
N_{zxxx} = -N_{zyyx},~\nonumber\\
N_{zxxy} = -N_{zyyy},~
N_{zxxz} = N_{zyyz},~
N_{zxyx} = -N_{zyyy},~
N_{zxyy} = N_{zyyx},~
N_{zxyz} = -N_{zyxz},~
N_{zxzx} = N_{zyzy},~\nonumber\\
N_{zxzy} = -N_{zyzx},~
N_{zxzz} = 0,~
N_{zyxx} = -N_{zyyy},~
N_{zyxy} = N_{zyyx},~
N_{zyzz} = 0,~
N_{zzxx} = N_{zzyy},~\nonumber\\
N_{zzxy} = -N_{zzyx},~
N_{zzxz} = 0,~
N_{zzyz} = 0,~
N_{zzzx} = 0,~
N_{zzzy} = 0.
\end{gather}
Using the notation of Eq.~(\ref{notation}), the reduced matrix can be obtained:
\begin{equation}
    N=\begin{pmatrix}
N_{11} & N_{12} & N_{13} & N_{14} & -N_{25} & N_{16} \\
N_{12} & N_{11} & N_{13} & -N_{14} & N_{25} & -N_{16} \\
N_{31} & N_{31} & N_{33} & 0 & 0 & 0 \\
N_{41} & -N_{41} & 0 & N_{44} & N_{45} & N_{52} \\
-N_{52} & N_{52} & 0 & -N_{45} & N_{44} & N_{41} \\
-N_{16} & N_{16} & 0 & N_{25} & N_{14} & N_{66}
\end{pmatrix},
\end{equation}
where $N_{66}=\frac{1}{2}(N_{11}-N_{12})$. For electrostriction and magnetostriction, this result aligns with Eq.~(92) of Ref. \cite{magnetostriction}.

For elasticity, we should further consider that the matrix is symmetric \cite{newnham} ($N=N^T$), and we have:
\begin{equation}
    N=\begin{pmatrix}
N_{11} & N_{12} & N_{13} & N_{14} & -N_{25} & 0 \\
N_{12} & N_{11} & N_{13} & -N_{14} & N_{25} & 0 \\
N_{13} & N_{13} & N_{33} & 0 & 0 & 0 \\
N_{14} & -N_{14} & 0 & N_{44} & 0 & N_{25} \\
-N_{25} & N_{25} & 0 & 0 & N_{44} & N_{14} \\
0 & 0 & 0 & N_{25} & N_{14} & \frac{1}{2}(N_{11} - N_{12})
\end{pmatrix},
\end{equation}
which aligns with Table 13.1 of Ref. \cite{newnham}.

\section{Discrete Model With Time Dependent Stimuli}\label{dis_gaussian}
Based on the discrete model, suppose that if the laser is a Gaussian pulse (Fig. \ref{fig:dynamic}e), the mathematical expression of its intensity is \cite{Gaussian_pulse}:
\begin{equation}
    a(t)=a_0~e^{-\frac{(t-t_0)^2}{\tau^2}}.
\end{equation}
Here, $a_0$ is the peak amplitude, $t_0$ is the center position, $\tau$ is the width parameter. Thus, the vector of laser is:
\begin{equation}
    \vec{E}_{lG}=a(t)\cdot\vec{E}_{l}.
\end{equation}
The equations about $\chi^{(2)}$ and $\vec{P}_{\text{m}}$ are:
\begin{align}
    \frac{d{\chi^{(2)}}}{dt}&=[a\sin^2(\theta+\beta)+b\cos^2(\theta+\beta)]\cdot\big(\vec{E}_{lG}\cdot \vec{E}_{lG}^T\big)\cdot{\chi^{(2)}}=a(t)^2\cdot A(\theta+\beta)\cdot{\chi^{(2)}},\nonumber\\
    \vec{P}_{\text{m}}(t)&=\chi^{(2)}(t)\cdot(\vec{E}_{lG})^{\otimes 2}=a(t)^2\cdot\chi^{(2)}(t)\cdot \big[R_z^T(\theta_n+\beta)\big]^{\otimes 2}\cdot(\vec{y}~')^{\otimes 2}.
\end{align}
And the process of Fig. \ref{fig:dynamic}e becomes:
\begin{enumerate}
    \item Step $1$: $\theta=\theta_1=0^\circ,~\vec{P}_{\text{m}~1}=a(0)^2\cdot\chi^{(2)}_{t=0}\cdot \big[R_z^T(\theta_1+\beta)\big]^{\otimes 2}\cdot(\vec{y}~')^{\otimes 2}$, during the time of $0\to t\to 2t_0 $:
    \begin{align}
    \frac{d\chi^{(2)}}{dt}= a(t)^2\cdot A(\theta_1+\beta)\cdot \chi^{(2)},~~
    \frac{d\vec{P}_{\text{m}}}{dt}=\left[ a(t)^2\cdot A(\theta_1+\beta)-\frac{4(t-t_0)}{\tau^2}\cdot\mathbb{I}_{2\otimes2}\right]\cdot \vec{P}_{\text{m}}=C(t)\cdot\vec{P}_{\text{m}}.
    \end{align}
The solutions are:
    \begin{align}
    \chi^{(2)}=\Phi(t,0)_1\cdot\chi^{(2)}_{t=0},~~\vec{P}_{\text{m}}=\Phi(t,0)_2\cdot\vec{P}_{\text{m}~1}.
    \end{align}
    where the $\Phi(t,0)_1$ and $\Phi(t,0)_2$ are:
\begin{align}
    \Phi(t,0)_1&=e^{\int_0^t  a(s)^2\cdot A(\theta_1+\beta)ds}=\text{exp}\bigg\{\frac{a_0^2~\tau}{2} \sqrt{\frac{\pi}{2}} 
\big[ 
\text{erf}( \frac{\sqrt{2}}{\tau} (t - t_0) ) 
+ 
\text{erf}( \frac{\sqrt{2}}{\tau} t_0)\big]\cdot A(\theta_1+\beta)\bigg\}=e^{k(t)\cdot A(\theta_1+\beta)},\nonumber\\
    \Phi(t,0)_2&=e^{\int_0^t C(s)ds}=e^{-\frac{2}{\tau^2}(t^2-2t_0t)}\cdot e^{k(t)\cdot A(\theta_1+\beta)}.
\end{align}
Here, we define:
\begin{align}
k(t)\equiv\frac{a_0^2~\tau}{2} \sqrt{\frac{\pi}{2}} 
\big[ 
\text{erf}( \frac{\sqrt{2}}{\tau} (t - t_0) ) 
+ 
\text{erf}( \frac{\sqrt{2}}{\tau} t_0)\big].
\end{align}
The ``$\text{erf}$'' is error function. When $t=2t_0$, the system changes to next step ($\theta=\theta_2$).

\item Step $2$: $\theta=\theta_2,~\vec{P}_{c~2}=a(0)^2\cdot\chi^{(2)}_{t=2t_0}\cdot \big[R_z^T(\theta_2+\beta)\big]^{\otimes 2}\cdot(\vec{y}~')^{\otimes 2}$, during the time of $2t_0\leq t\leq 4t_0$.
    \begin{align}
    \frac{d\chi^{(2)}}{dt}= a(t)^2\cdot A(\theta_2+\beta)\cdot \chi^{(2)},~~
    \frac{d\vec{P}_{\text{m}}}{dt}=\left[a(t)^2\cdot A(\theta_2+\beta)-\frac{4(t-t_0)}{\tau^2}\cdot\mathbb{I}_{2\otimes2}\right]\cdot \vec{P}_{\text{m}}=C'(t)\cdot\vec{P}_{\text{m}}.
    \end{align}
    The solutions are:
    \begin{align}
    \chi^{(2)}&=e^{ k(t) \cdot  A(\theta_2+\beta)}\cdot\chi^{(2)}_{t=2t_0}=e^{ k(t)\cdot A(\theta_2+\beta)}\cdot e^{k(2t_0)\cdot A(\theta_2+\beta)}\cdot\chi^{(2)}_{t=0},\nonumber\\
    \vec{P}_{\text{m}}&= e^{-\frac{2}{\tau^2}(t^2-2t_0t)}\cdot e^{k(t)\cdot A(\theta_2+\beta)}\cdot\vec{P}_{\text{m}~2},\nonumber\\
    &=e^{-\frac{2}{\tau^2}(t^2-2t_0t)}\cdot e^{k(t)\cdot A(\theta_2+\beta)}\cdot e^{k(2t_0)\cdot A(\theta_1+\beta)}\cdot\chi^{(2)}_{t=0}\cdot \big[R_z^T(\theta_2+\beta)\big]^{\otimes 2}\cdot(\vec{y}~')^{\otimes 2}.
    \end{align}
    Thus, for the step $n$:
    \begin{align}
    \vec{P}_{\text{m}}&=a(0)^2\cdot[e^{-\frac{2}{\tau^2}(t^2-2t_0t)}\cdot e^{k(t)\cdot A(\theta_n+\beta)}]\cdot\left[ \prod_{i=0}^{n-1} e^{k(2t_0)\cdot A(\theta_i+\beta)}\right]\cdot\chi^{(2)}_{t=0}\cdot \big[R_z^T(\theta_n+\beta)\big]^{\otimes 2}\cdot(\vec{y}~')^{\otimes 2}.
    \end{align}
\end{enumerate}
If we using the value of $\vec{P}_{\text{m}}$ at $t_0$ to represent the test data, the $I_{\text{SHG}}$ at $\theta_n$ is:
\begin{equation}
    I_{\text{SHG}}(\theta_n)=\left|\vec{E}_l(\theta_n)^T\cdot \vec{P}_{\text{m}}(\theta_n,t_0)\right|^2.
\end{equation}
Its scatters are shown in Fig.~\ref{fig:dynamic}h.

\end{document}